\documentclass{article}
\usepackage{amsmath,graphicx,url,amssymb,latexsym,epsfig,tabularx,setspace,verbatim,rotating,threeparttable}
\usepackage{natbib}
\usepackage[hmargin=1.5in,vmargin=1.5in]{geometry}

\begin{document}
\setcounter{secnumdepth}{4}
\begin{titlepage}
\centering
\textbf{\Large A Model of Habitability Within the Milky Way Galaxy }\\
M. G. Gowanlock$^{1,2,3}$, D. R. Patton$^{1}$, and S. M. McConnell$^{2}$ \\
\vspace{2cm}
$^{1}$Department of Physics \& Astronomy, Trent University, Peterborough, ON, Canada

$^{2}$Department of Computing \& Information Systems, Trent University, Peterborough, ON, Canada

$^{3}$Department of Information \& Computer Sciences and NASA Astrobiology Institute, University of Hawaii-Manoa, Honolulu, HI, U.S.A.

\vspace{1cm}
\raggedright
Corresponding author:\\
Michael Gowanlock\\
Department of Information \& Computer Sciences, University of Hawaii, POST 310, 1680 East-West Road, Honolulu, HI, 96822, USA\\
E-mail: gowanloc@hawaii.edu
\end{titlepage}

\singlespace

\begin{abstract}
We present a model of the Galactic Habitable Zone (GHZ), described in
terms of the spatial and temporal dimensions of the Galaxy that may
favour the development of complex life.  The Milky Way galaxy is modelled
using a computational approach by populating stars and their planetary
systems on an individual basis using Monte-Carlo methods.  We begin with
well-established properties of the disk of the Milky Way, such as the stellar number
density distribution, the initial mass function, the star formation
history, and the metallicity gradient as a function of radial position
and time.  We vary some of these properties, creating four models to test the sensitivity of our assumptions. To assess habitability on the Galactic scale, we model
supernova rates, planet formation, and the time required for complex
life to evolve.  Our study improves on other literature on the GHZ by populating stars on an individual basis and by modelling SNII and SNIa sterilizations by selecting their progenitors from within this preexisting stellar population.  Furthermore, we consider habitability on tidally locked and non-tidally locked planets separately, and study habitability as a function of height above and below the Galactic midplane.  In the model that most accurately reproduces the properties of the Galaxy, the results indicate
that an individual SNIa is $\sim$5.6$\times$ more lethal than an individual SNII on average.  In addition, we predict
that $\sim$1.2\% of all stars host a planet that may have been capable
of supporting complex life at some point in the history of the Galaxy.  Of those stars with a habitable planet,
$\sim$75\% of planets are predicted to be in a tidally locked configuration with their host star.  The majority of these planets that may
support complex life are found towards the inner Galaxy, distributed
within, and significantly above and below, the Galactic midplane.
\end{abstract}


\section{Introduction}
\par
Studies of habitability on the Galactic scale are gaining attention as extrasolar planetary searches and advanced models of the evolution of the Milky Way galaxy improve our understanding of the prerequisites for life.  The GHZ is the area in the Galaxy that may favour the emergence of complex life.  The inner boundary of the GHZ is thought to be defined by hazards to planetary biospheres, and the outer boundary is set by the minimum amount of metallicity required for planet formation \citep{2001Icar..152..185G}.  Given these factors, the GHZ is constrained by the morphology, stellar populations, and chemical evolution of the Milky Way.  The GHZ is defined as a concept analogous to the Circumstellar Habitable Zone (HZ), which is the region around a star where water can remain in a liquid state on a planet's surface.

\par
To assess habitability on the Galactic scale, we model planet formation, supernova rates, and the time required for complex life to evolve.  Estimates in predicting the number of habitable planets in the Galaxy are in their infancy at present.  Candidate habitable planets have been identified in the Gliese 581 system \citep{2007A&A...476.1373S,2010ApJ...723..954V}, and in the Kepler list of candidate exoplanets \citep{2011arXiv1102.0541B}. Models can provide insight into the number and distribution of habitable planets that are likely to exist in the Galaxy.  Prior studies suggest that there is a correlation between high-metallicity environments and planet formation \citep{2004ApJ...616..567I,2004A&A...415.1153S,2005ApJ...622.1102F,2007ApJ...669.1220G,2009ApJ...697..544S,2010PASP..122..905J}.  The metallicity gradient found in the disk of the Galaxy has been measured, and suggests that the number of planets depends on radial position.  Regions greatly affected by the detrimental impacts of supernovae (SNe) are defined by proximity to these transient events.  Therefore, the stellar density and star formation history (SFH) directly influence the frequency at which a planet's biosphere is negatively affected.  Finally, in combination with the periods that SNe are expected to sterilize life in surrounding solar systems, enough time must elapse such that complex life can evolve.
\par
Research on the GHZ by \citet{lineweaver-GHZ} suggests that the GHZ is an annular region that expands with time.  However, \citet{2008SSRv..135..313P} argues that the entire disk of the Galaxy may be suitable for complex life.  From our work we predict that at the present time, the entire disk of the Galaxy is hospitable to complex life; however, the regions that permit the majority of complex life are contrary to other predictions in the field.
\par
Predicting where complex life might be favoured in the Galaxy follows from making assumptions about the prerequisites for life.  Whether or not life exists beyond the context of our solar system cannot be validated at present.  Our definitions of habitability in the Galaxy do not imply any claims of inhabitance.
\par
We model the GHZ with the factors described above.  We begin in \S~\ref{model} by outlining the construction of our model Milky Way galaxy.  In \S~\ref{model_habitability_MWG} we describe the factors that influence habitability on the Galactic scale.  These factors include: the dangers caused by SN sterilizations, how the metallicity gradient influences planet formation, and the timescale requirements for the emergence of biologically complex life.  Results and a comparison with other studies are discussed in~\S~\ref{results}.  Finally, in~\S~\ref{conclusion} we present the conclusions drawn from our study of habitability in the Galaxy.

\section{A Model of the Milky Way Galaxy}\label{model}
\par
In this section we describe our model of the Milky Way.  We assign the birth date and metallicity of each star in a self-consistent way by matching the SFH and the metallicity gradient of \citet{2006MNRAS.366..899N}.  To assess the sensitivity of our model, we vary the initial mass function (IMF) and the stellar number density distributions.  Varying the stellar number density distributions creates differing profiles of the star formation rate (SFR) within the Galactic disk.  We do not try to match the number density of \citet{2006MNRAS.366..899N}, as their model does not describe the vertical component of the disk. Rather, we match the number density distributions of \citet{2006ima..book.....C} and \citet{2008ApJ...673..864J}, to permit us to study the Galaxy as a function of distance above the midplane and radial distance from the Galactic centre.  This approach yields a three-dimensional model that allows us to study habitability as a function of radial distance and height above the midplane.

\subsection{Stellar Mass and Main Sequence Lifetime}\label{stellar_mass}
\par
This subsection outlines how a mass and main sequence lifetime are assigned to each star in our model Milky Way.  We assign stellar masses using a Monte Carlo technique such that the resulting distribution follows a power-law and matches an IMF with $\alpha=2.35$ \citep{1955ApJ...121..161S}.  The maximum and minimum stellar masses for main sequence stars are defined as 100 M$_\odot$ and 0.08 M$_\odot$ respectively \citep{1993MNRAS.262..545K}.
\par
This IMF is consistent with observations of star-forming regions; however, there are uncertainties \citep{1999ApJS..123....3L}, and other IMFs have been proposed. \citet{1979ApJS...41..513M}, \citet{2001MNRAS.322..231K}, and \citet{2003PASP..115..763C} suggest that at subsolar masses, $\alpha$ should be lower than 2.35, flattening the IMF.  \citet{2006MNRAS.366..899N} test both the Salpeter and Chabrier IMF (flattening at subsolar masses) in their model of the Milky Way and find that both are in agreement with observations.  One of the IMFs we utilize is the Salpeter IMF to remain consistent with the model Galaxy by \citet{2006MNRAS.366..899N}.
\par
To test the impact of the IMF flattening, in comparison to the Salpeter IMF at subsolar masses, we implement the IMF given by \citet{2001MNRAS.322..231K}, which is described as a two part power law function.  The value $\alpha=1.3$ when 0.08 $\leq$ M $<$ 0.5, and $\alpha=2.3$ when M $\geq$ 0.5.  The steep slope at subsolar masses in the Salpeter IMF suggests that the Galaxy is populated with many more low mass stars than the Kroupa IMF.

\par
The mass corresponds to a main sequence lifetime defined by \citet{1994sipp.book.....H}:
\begin{equation}T_L=T_{L\odot}\left(\frac{m_\odot}{m}\right)^{2.5},\label{eqn:stellar_lifetime}\end{equation} where $m_\odot=1$ is the Sun's mass, $T_{L\odot}=11$ is the Sun's main sequence lifetime in Gyr \citep{1993ApJ...418..457S}, and $m$ is the star's mass determined from the Salpeter or Kroupa IMF.

\subsection{Star Formation History}
\par
This subsection outlines when stars form in our model of the Milky Way.  We adopt the SFH given by \citet{2006MNRAS.366..899N}, who model the SFH in the Galactic disk over time.  Their model yields an inside-out formation of the Galaxy, with an early burst of star formation in the inner Galaxy that declines over time. Close inspection of Figure~9 in \citet{2006MNRAS.366..899N} reveals that there are fluctuations in the SFH, as natural processes such as the passage of spiral waves \citep{2000MNRAS.316..605H} and close encounters with the Magellanic Clouds \citep{2000A&A...358..869R} trigger star formation.  These sophisticated mechanisms that cause fluctuations in star formation are not modelled in the present study. Fitting the curves and interpolating between them in Figure~6 (SFR vs. radius) from \citet{2006MNRAS.366..899N} establishes when the stars form in our model.  We recognize that other SFHs have been reported \citep{2009MNRAS.397.1286A,2009ApJ...696..668F}, although we do not model them in this study. The SFH adopted here varies by radial position (R), but is independent of the vertical (z) position in the Galaxy.

\subsection{Distribution of Stars Within the Model Milky Way Galaxy}\label{stellar_distribution}
\par
We utilize the distribution of stars found by \citet{2006ima..book.....C} and \citet{2008ApJ...673..864J} to determine where the stars form in our model of the Milky Way.  We vary the number density distributions to test how sensitive the model is to differing profiles of this parameter.

\subsubsection{Distribution of Stars by \citet{2006ima..book.....C}}

Their formula describes the number density of stars as a function of distance above the midplane and radial distance from the Galactic centre, as follows:  \begin{equation}n(z,R)=n_0(e^{-z/z_{thin}}+ 0.085e^{-z/z_{thick}})e^{-R/h_R}.\label{eqn:stellar_density}\end{equation} The quantity $n$ is the number of stars per unit volume (pc$^3$). The coordinate $z$ is the vertical height above the midplane of the Galaxy, and $R$ is the radial distance from the Galactic centre. The constant $h_R=2.25$ kpc denotes the radial disk scale length. The thin disk lies within the central plane of the Galactic disk with a scale height of $z_{thin}\simeq$ 350 pc, and the thick disk has a vertical scale height of $z_{thick}\simeq$1000 pc.  The quantity $n_0$ (number of stars per pc$^{3}$) is the normalization that we will select below.
\par
We normalize the distribution of stars based on an estimate of the total disk mass in the Milky Way.  The disk mass estimate of \citet{2008gady.book.....B} yields $4.2\times10^{10}$M$_{\odot}$ in disk stars ($0.3\times10^{10}$M$_{\odot}$ has been subtracted, as it corresponds to gas mass).  We are able to match this disk mass using our Salpeter IMF described in $\S$~\ref{stellar_mass} with a value of $n_0$=5.502 stars pc$^{-3}$.   This normalization yields $\sim$150 billion stars in the disk of the Galaxy.  With regards to the Kroupa IMF, we normalize $n_0$=1.957 stars pc$^{-3}$, yielding $\sim$50 billion stars in the disk of the Galaxy.
\par
To ensure our distribution is appropriate, we compare our estimates to the number density of the local neighbourhood.  For the Salpeter IMF, this normalization overpredicts by a factor of $\sim$1.4 the local number density reported by \citet{2002AJ....124.2721R}.  Other disk mass estimates \citep[for example,][]{2006ima..book.....C} are even higher than that of \citet{2008gady.book.....B}, and would overpredict the local density to a greater degree than the normalization utilized here.  Note that the 40\% discrepancy refers only to the local neighbourhood, and would be unacceptable if it applied to the entire disk.  For the Kroupa IMF, the local number density is $\sim$50\% of that found by \citet{2002AJ....124.2721R}.  The Salpeter and Kroupa IMFs, paired with the stellar number density distribution of \citet{2006ima..book.....C} are referred to as Model~1 and Model~2 respectively.

\subsubsection{Distribution of Stars by \citet{2008ApJ...673..864J}}

Their formula describes the number density of stars as a function of distance above the midplane and radial distance from the Galactic centre, as follows:
\begin{equation}\rho_D(R,Z)=\rho_D(R,Z;L_1,H_1)+f\rho_D(R,Z;L_2,H_2),\label{eqn:stellar_density2}\end{equation}
where
\begin{equation}\rho_D(R,Z;L,H)=\rho_D(R_\odot,0)e^{R_\odot/L}\times e^{\big(-\frac{R}{L}-\frac{Z+Z_\odot}{H}\big)}.\label{eqn:stellar_density2a}\end{equation}
The quantity $\rho_D$ is the number of stars per unit volume (pc$^3$).  The coordinate $Z$ is the vertical height above the midplane of the Galaxy, and $R$ is the radial distance from the Galactic centre.  We utilize the values $H1$ = 300 pc, $L1$ = 2600 pc, $H2$ = 900 pc, $L2$ = 3600 pc, and $f$ = 0.12, corresponding to the thin disk scale height and length, the thick disk scale height and length and the thick-to-thin disk density normalization.
\par
Using the Salpeter IMF, we are able to match our disk mass ($4.2\times10^{10}$M$_{\odot}$) by normalizing $\rho_D(R_\odot,0)$=0.237 stars pc$^{-3}$.  We normalize $\rho_D(R_\odot,0)$=0.084 stars pc$^{-3}$, corresponding to the Kroupa IMF.
\par
The normalization for the Salpeter IMF overpredicts the local number density reported by \citet{2002AJ....124.2721R} by a factor of $\sim$2.  When implementing the Kroupa IMF, the local number density is $\sim$70\% of that found by \citet{2002AJ....124.2721R}.  The Salpeter and Kroupa IMFs, paired with the stellar number density distribution of \citet{2008ApJ...673..864J} are referred to as Model~3 and Model~4 respectively.

\subsubsection{The $R$ and $z$ Dependence of the Number Density Distributions}
\par
Our model Galaxy has both $R$ and $z$ dimensions.  The vertical motions of stars passing through the midplane, which are not included in this study, would serve to reduce (but not remove) trends in the $z$ dimension; however, we do not expect trends in habitability to be significantly diminished.  Stars spend different amounts of time above and below the midplane, as all stars do not bob up and down to the same extent.  For example, the Sun is expected to remain within 70 pc above or below the midplane \citep{2005ApJ...626..844G}.  Therefore, the Sun is well within a scale height of the midplane throughout its vertical motions.  We do not believe that the absence of vertical and similarly, radial motions in our model to substantially affect trends as a function of $z$ or $R$.
\par
Our model of the Milky Way ignores binary star systems, spiral arms, the orbits of stars, the production of stars in clusters, and the accretion of satellite galaxies over time.  These are observable properties of the Galaxy that could have an impact on habitability, but are not implemented in the present study.  While other studies use more sophisticated simulations of the Milky Way, we populate stars on an individual basis.  Our approach is reasonable, since it reproduces the global properties of the Galaxy.
\par
The SFR produced in our model as a function of radial distance and time is shown in the upper (Model~1) and middle (Model~4) panels of Figure~\ref{fig:SFR-time}.  As a result of the inside-out nature of Galaxy formation, the inner Galaxy has an early burst of star formation that decreases with time.  The SFR around the solar neighbourhood has remained constant over the past few Gyr, and the SFR has been increasing in the outskirts.

\subsection{Metallicity}
Through the continual births and deaths of generations of stars, the abundance of metals increases in the Galaxy over time.  Metals are those elements heavier than helium, and are important as they are the building blocks of terrestrial planets.  Stars that form at a later date have a greater chance of having terrestrial planets than earlier generations of stars.  We use metallicity as a proxy for planet formation, which is consistent with other literature on the subject \citep{2004ApJ...616..567I,2004A&A...415.1153S,2005ApJ...622.1102F,2007ApJ...669.1220G,2009ApJ...697..544S,2010PASP..122..905J}. A metallicity profile is used to assign an appropriate metallicity in terms of [Fe/H] abundance to each star in the model\footnote{If a star is formed before 2 Gyr, it is assigned the metallicity of a star that formed at 2 Gyr.}.  There are a number of models that map the evolution of the metallicity gradient in the Galaxy.  We utilize the metallicity profile in Figure~11 of \citet{2006MNRAS.366..899N}.  We use a fixed metallicity for stars born at a given radial position and time, which is consistent with \citet{2006MNRAS.366..899N}.  The average metallicity of the stars produced in terms of radial position and time in our model is shown in the lower panel of Figure~\ref{fig:SFR-time}, and the distribution of metallicities for all stars that form at the solar radius (R$\sim$8 kpc) is shown in Figure~\ref{fig:metallicity_dist_8kpc}.  The model by \citet{2006MNRAS.366..899N} is consistent with literature regarding the metallicity distribution of stars within the solar neighbourhood. At this location, \citet{2008ApJ...684..691R} find that 35\% of local stars have Z$>$Z$_\odot$, whereas 41\% of the stars in our model have Z$>$Z$_\odot$.  Metallicity varies as a function of $R$ and time in our model, whereas a more complex model could also depend on $z$.  Metallicity trends as a function of $z$ may be minimal, as studies suggest that there is little variation in the vertical gradient within the thick disk \citep{1995AJ....109.1095G}.

\par
The chemical evolution of the Galaxy is affected by radial mixing.  Radial mixing has the effect of modestly flattening the metallicity gradient across the Galactic disk; however, the gradient is steep enough such that planet formation is noticeably varied throughout the disk of the Galaxy.  Therefore, like \citet{2006MNRAS.366..899N} we do not account for radial mixing in our model, as this mechanism would not be responsible for significantly changing the observed trends in the results.

\subsection{Varying Models of the Milky Way}
We vary the number density distribution and IMF to see how sensitive our model is to different estimates of the properties of the Milky Way. Table~\ref{ex:model_description} gives an overview of our models of the Galaxy.  We focus on Models 1 and 4 in our results, as they are the most consistent with the  number density of the solar neighbourhood as reported by \citet{2002AJ....124.2721R}.  Model~1 overpredicts the local stellar number density by a factor of $\sim$1.4, and Model~4 underestimates the local stellar number density, and is $\sim$70\% of the value found by \citet{2002AJ....124.2721R}.  These two models bracket the stellar number density in the solar neighbourhood.
\subsection{Model Implementation and Properties of the Galaxy}
\par
Our model of the Milky Way disk has a radius of 15 kpc.  We present our findings for the 2.5-15 kpc region of the disk, as the bulge overlaps the disk at $R\lesssim$2.5 kpc.  We take advantage of the azimuthal symmetry within the Milky Way by modelling only a 1$^\circ$ sector of the disk.  To eliminate uncertainties with this approach we introduce periodic boundary conditions when processing SNe sterilizations.  Stars are processed on an individual basis; when processing SNe sterilizations, we keep track of the times in which SNe are sufficiently close to each star.  We iterate through all of the stars in the model without using a brute-force approach.

\section{A Model of Habitability Within the Milky Way Galaxy}\label{model_habitability_MWG}
Various properties were reviewed in \S~\ref{model} that are used to construct a model of the Milky Way for the purposes of assessing habitability on the Galactic scale.  In the following subsections we outline some prerequisites and conditions for land-based animal life.  These prerequisites are used to assess if there is a position and time in the Galaxy that favours the emergence of complex life.
\subsection{The Effects of Supernovae on a Planet's Biosphere}\label{SN}
\par
Cosmic rays, $\gamma$-rays and X-rays emitted by a supernova can lead to the sterilization of life in surrounding solar systems
\citep[for an overview of astrophysical radiation sources see][]{2011AsBio..11..343M}. The depletion of ozone in a planetary atmosphere caused by a SN exposes a planet's surface to more radiation from its host star \citep{2003ApJ...585.1169G}.  We model Type II SNe (SNII) and Type Ia SNe (SNIa) independently to reflect the differences in their formation rates, luminosities and consequent sterilization distances.  SNe that are not SNIa or SNII are ignored, as they are responsible for only a small fraction of SNe that occur in all types of galaxies \citep{1999A&A...351..459C}.  This method differs from the work of \citet{lineweaver-GHZ} and \citet{2008SSRv..135..313P}.  Our approach treats both SN types separately and permits a range of sterilization distances for SNII and SNIa.  We model SNe in a self-consistent way by selecting SN progenitors from within our pre-existing stellar distribution, whereas \citet{lineweaver-GHZ} and \citet{2008SSRv..135..313P} use a time-integrated supernova danger factor which depends on the supernova rate, but is normalized in an arbitrary fashion.
\subsubsection{Supernovae Sterilizations in Other Studies}
\par
The previous work of \citet{lineweaver-GHZ} and \citet{2008SSRv..135..313P} use the SNe rate in the Galaxy to determine the probability that life on a planet survives SNe explosions across the Galactic disk.  They normalize the danger posed by SNe to the time integrated SN rate of the solar neighbourhood. They suggest that if the time integrated rate is 4$\times$ that of the solar neighbourhood, the probability of surviving SNe explosions is 0, and if the time integrated rate is $<$0.5 that of the solar neighbourhood, the probability of surviving SNe explosions is 1.
\par
Whether a planet survives a SNe explosion is linked to the SNe rate; however, the time in which a SN occurs in a planet's history is also a major factor.  For example, if a planet is irradiated by multiple SN explosions early in its history, the sterilizations should not preclude it from being habitable at later dates.  In an analogous scenario, the late heavy bombardment has not prevented the Earth from attaining complex life.  We test to see when each individual planet was irradiated by SN events, which permits us to more accurately determine if a planet has the potential to host complex life.  This is described in Section~\ref{life}.  Furthermore, we do not model a SN rate; however, if we were to test the rate at various positions, ours would depend on $R$ and $z$, whereas the related work is only dependant on $R$.

\subsubsection{Supernovae Sterilizations in the Present Study}
\par
Stars with a main sequence mass greater than $\approx$8~$M_\odot$ become SNII at the end of their lifetimes \citep{1984ApJ...277..361K}.  In our model, all stars with a main sequence mass $>$8~$M_\odot$ become SNII after their respective main sequence lifetime expires.  We do not account for the delay between the end of the main sequence lifetime of a star and the time before it explodes as a SN; however, the time delay before a SN occurs is relatively small, as a large majority of a star's lifespan is spent on the main sequence.  Models 1 and 3 (Salpeter IMF) yield a Galactic SNII rate over the past Gyr of 1.98-2.40$\times$10$^{-2}$ yr$^{-1}$, consistent with the SNII rate of \citet{1993A&A...273..383C}, who find a Galactic SNII rate of 0.4-2$\times$10$^{-2}$yr$^{-1}$. Models 2 and 4 (Kroupa IMF) yield a Galactic SNII rate over the past Gyr of 3.70-4.40$\times$10$^{-2}$ yr$^{-1}$, consistent with the SNII rate by a factor of $\sim$2 of \citet{1993A&A...273..383C}.  The average SNII rate in \citet{1993A&A...273..383C} is calculated by using the luminosity of the Galaxy.
\par
Theoretical estimates by \citet{1992ApJ...386..197T} predict a SNII rate of 1.96-3.35$\times$10$^{-2}$ yr$^{-1}$, which is closer to the predicted values in the present study.  The Kroupa IMF leads to more high mass stars than the Salpeter IMF; therefore, the Kroupa IMF yields more SNII.~\citet{1993A&A...273..383C} note that six Milky Way SNe have been observed within the last millennium, but many additional Galactic SNe have gone undetected due to absorption by dust in the disk. Predictions of the true SNe rate vary considerably in the literature.
\par
The type of SNIa focused on in our model is the single degenerate case, wherein a white dwarf in a binary system grows in mass due to accretion from a binary companion.  When the Chandrasekhar limit of $1.4M_\odot$ is reached, the star explodes as a result of insufficient electron degeneracy pressure. \citet{2008ApJ...683L..25P} estimate that $\sim$1\% of all white dwarfs become SNIa, independent of mass.  In our model, all stars that are white dwarf candidates have main sequence masses of $0.08~M_\odot<M<8~M_\odot$, and are assigned a 1\% chance of becoming SNIa after their main sequence lifetime has expired.  The detonation delay times implicit in the \citet{2008ApJ...683L..25P} model are consistent with the delay time estimates of \citet{2009ApJ...707...74R}.  Models 1 and 3 (Salpeter IMF) yield a SNIa rate of 4.0$\times$10$^{-3}$ yr$^{-1}$ over the past Gyr, consistent within a factor of $\sim$1.5 of the SNIa rate of \citet{2010ApJ...710.1310M}, who find a Galactic SNIa rate of 2.25-2.9$\times$10$^{-3}$ yr$^{-1}$.  Models 2 and 4 (Kroupa IMF) yield a SNIa rate of 6.6$\times$10$^{-3}$ yr$^{-1}$ over the past Gyr, consistent within a factor of $\sim$2.5 of the SNIa rate of \citet{2010ApJ...710.1310M}.  The rate of SNIa is lower than that of SNII because the majority of SNIa candidates are very low mass stars and long timescales are required before they evolve off the main sequence.  We recognize that the double degenerate channel is responsible for some SNIa \citep{2010ApJ...710.1310M}; however, observations indicate that the single degenerate channel is dominant~\citep{2007NewAR..51..524P}.
\par
\citet{2003ApJ...585.1169G} find that at a distance of $<8$ pc, a SNII will deplete the ozone in a planet's atmosphere to the point where the UV flux received from the local host star will have a sterilizing effect on any land-inhabiting life that exists on the planet.  We assume that the sterilization distance outlined by \citet{2003ApJ...585.1169G} refers to an average SNII and is just sufficient to sterilize life within 8 pc.  We scale the sterilization distances of all SNII and SNIa accordingly. \citet{2002AJ....123..745R} address the distribution of absolute magnitudes for SNII.  Considered in the study are 54 observed SN events in external galaxies, and the mean absolute magnitudes in the B band ($M_B$) are found for each of the SNII subclasses (SNII-L, SNII-P and SNIIn).  We fit the data across the SNII subclasses to determine the distribution of absolute magnitudes of SNII.  In addition, we use the distribution of SNIa absolute magnitudes reported by \citet{2006ApJ...645..488W}, which is based on 109 SNIa with a mean $M_B$ of $-$19.34.

\par
Given these distributions of SNe absolute magnitudes, we compute the sterilization distances for each SNII and SNIa using the following:
\begin{equation}d_{SN}=8~{\rm pc}\times\sqrt{10^{-0.4(M_{SN}-M_{std})}},\end{equation}
where $d_{SN}$ is the resulting sterilization distance for a given SN, $M_{SN}$ is its absolute magnitude, and $M_{std}=-17.505$, which is the absolute magnitude of the average SNII that we assume is just sufficient to sterilize life within 8 pc.

Our resulting distribution of sterilization distances, corresponding to the distribution of absolute magnitudes of SNII and SNIa, are shown in Figure~\ref{fig:sterilization_distances}.  SNIa are more luminous than SNII on average, and therefore are expected to sterilize planets at a greater distance than SNII.

\subsection{Metallicity and Planet Formation}\label{metallicity}
\par
Predicting a region that may favour complex life in the Galaxy would be simplified given a census of habitable planets in the Milky Way.  The Kepler mission will eventually yield an estimate of this nature.  Of the 1202 Kepler exoplanet candidates discussed in \citet{2011arXiv1102.0541B}, 6 exist in the HZ that are less than twice the size of the Earth.  In the meantime, we use the metallicity-planet correlation \citep{2004ApJ...616..567I,2004A&A...415.1153S,2005ApJ...622.1102F,2007ApJ...669.1220G,2009ApJ...697..544S,2010PASP..122..905J} in combination with our model of metallicity within the Galaxy and a model of solar system formation to predict where habitable planets will be found.  This subsection describes how habitable planets are assigned to individual stars in our model Milky Way.

\subsubsection{Planet Formation in Other Studies}
\par
\citet{lineweaver-GHZ} calculate the probability of terrestrial planet formation in their model of the GHZ as influenced by \citet{2001Icar..151..307L}. \citet{2001Icar..151..307L} suggests that too little metallicity will inhibit the formation of planets and too much metallicity will produce giant planets that may hinder the formation or survival of habitable planets.  Giant planets may reduce the number of habitable planets through the following processes: 1) the planet will be ejected out of the solar system or into its host star through the process of gravitational interaction, or 2) the protoplanetary disk will be consumed by the migrating giant planet, such that there is not enough material to form terrestrial planets, or the migrating giant planet accretes habitable planets. \citet{lineweaver-GHZ} find that the potential of a star to harbour Earth-like planets increases linearly until a high metallicity abundance suggests the formation of massive planets.  As the probability of having massive planets around a host star increases, the probability that Earth-like planets survive decreases.

\par
Conversely, \citet{2008SSRv..135..313P} argues for a fixed planet formation probability of 40\% above 0.1~$Z_\odot$.  This is justified by the statement that Earth-like planets may be common in low-metallicity environments.  The metallicity integrated probability of forming an Earth-like planet is the same as in the work by \citet{lineweaver-GHZ}. \citet{2008SSRv..135..313P} estimates the probability of forming Hot Jupiters using the metallicity-planet correlation quantified by \citet{2005ApJ...622.1102F}. This work is also utilized in our model.  The probability of having a star with a habitable planet without a Hot Jupiter that would destroy it extends to lower and higher metallicity ranges than that of \citet{lineweaver-GHZ}.  For comparative purposes, the probability of forming an Earth-mass planet in the model by \citet{2008SSRv..135..313P} is much higher than the probability adopted in our model.

\subsubsection{Planet Formation in the Present Study}
\par
The probability of forming a habitable planet or a migrating massive planet (hereafter referred to as a Hot Jupiter) is estimated using the metallicity of each host star in the model.  We use data from simulations by \citet{2005ApJ...626.1045I} as well as research by \citet{2005ApJ...622.1102F} to estimate the number of habitable planets and Hot Jupiters in our model.
\par
The probability of forming a gas giant planet around a star is utilized to estimate the probability of forming a habitable planet or Hot Jupiter. \citet{2005ApJ...622.1102F} find that the probability of forming a gas giant planet is
\begin{equation} P(planet)=0.03\times10^{2.0[Fe/H]}.\label{eqn:form_planet}\end{equation}
\citet{2004A&A...415.1153S} also quantify the planet-metallicity correlation, and find that a flat metallicity tail exists, wherein the probability of forming a planet is constant (3\%) at sub-solar metallicities ([Fe/H]$<$0).

\subsubsection*{The Probability of Forming a Hot Jupiter}
We assume that the probability of forming a Hot Jupiter is 3\% at sub-solar metallicities and that the probability of forming a Hot Jupiter at [Fe/H]$>$0 is given by Equation~\ref{eqn:form_planet}. \citet{2008SSRv..135..313P} estimates the probability of forming a Hot Jupiter using a similar method.

\subsubsection*{The Probability of Forming a Habitable Planet}
\par
To link the metallicity-planet correlation of gas giant planets to habitable planets, we employ a model of solar system formation that contains both habitable planets and planets that would be detectable by the radial velocity technique (short period gas giants).  The model of \citet{2005ApJ...626.1045I} produces short period gas giants and habitable planets in roughly equal numbers.  We note that in the work of \citet{2005ApJ...622.1102F}, nearly all of the planets studied satisfy the radial velocity criteria of \citet{2004ApJ...616..567I}.   In particular, in Figure~2 of \citet{2005ApJ...626.1045I} the distributions of the semimajor axis and mass of planets around stars of 0.2, 0.4, 0.6, 1.0 and 1.5 $M_\odot$ are shown.  Note that we assume a planet is considered habitable if it exists in the HZ around its host star (allowing liquid water on the planetary surface), and has $0.1~M_\oplus<M_p<10~M_\oplus$, where $M_p$ is the mass of the planet \citep{1993Icar..101..108K}.  A planet is within the HZ if its semimajor axis ($a$) lies within the range $0.8-1.5\times(M_*/M_\odot)^2$ AU \citep{1993Icar..101..108K}, where $M_*$ is the mass of the star.

\par
To associate the metallicity-planet correlation for gas giant planets to habitable planets, we find the ratios of HZ planets with $0.1~M_\oplus<M_p<10~M_\oplus$ to planets detectable by radial velocity in \citet{2005ApJ...626.1045I}\footnote{Data obtained through private communication with S. Ida.}.  A planet is deemed to be detectable by the radial velocity method if $M_p\gtrsim100(a/1~AU)^{1/2}~M_{\oplus}$ and $a\lesssim3$ AU, as discussed in \citet{2004ApJ...616..567I}.  The ratios are roughly 1:1.\footnote{The ratios are not explicitly shown as the data belongs to \citet{2005ApJ...626.1045I}.}  We scale the probability of forming a habitable planet by these ratios, which are binned with the following ranges of $M_*$: 0.08-0.5, 0.5-0.8, 0.8-1.25, and 1.25-$\sim$1.5.  The probability of forming a habitable planet with [Fe/H]$>$0 is given by Equation~\ref{eqn:form_planet} scaled by the ratios described above.  At sub-solar metallicities, the probability of forming a habitable planet is 3\% and scaled by the same method. \citet{2005ApJ...626.1045I} permit only one major planet around each star in their model.  Therefore, a maximum of one habitable planet can exist around any star in our model.  This study does not advance the notion that the stars in extra-solar systems are expected to contain a single planet.

\par
Given the probability of forming a habitable planet or Hot Jupiter, if a host star has a habitable planet and a Hot Jupiter, we assume that the habitable planet does not survive.  The probability of a star having a habitable planet without a Hot Jupiter that would destroy it is modelled as a result of the metallicity abundance of its host star. The study by~\cite{2010Sci...330..653H} suggests that 23\% of stars harbour a short period Earth-mass (0.5-2 $M_\oplus$) planet.  Although this finding does not suggest that all of these planets exist in the HZ of their host stars, we find that $\sim$5\% of all stars host a habitable planet ($0.1~M_\oplus<M_p<10~M_\oplus$) in the HZ of their host star across the disk of the Galaxy.  Our study may underestimate the number of habitable planets that exist in the Milky Way.  The locations of our habitable planets populated over all epochs in Models~1 and 4 are shown in Figure~\ref{fig:all_planets}.  The inner Galaxy hosts the greatest number of habitable planets due to the high metallicity abundance and stellar density in the region.  Note that these are the locations of all the habitable planets that have survived possible Hot Jupiter migrations, before determining if they have survived SNe sterilizations.  Therefore, all of these planets will not be suitable for advanced biological life; however, a large portion may be suitable for microbial life.

\subsection{Tidal Locking of Planets around Stars}
\par
The distribution of the number and types of planets around stars of various masses is not fully understood at present.  However, low-mass stars are not precluded from having Earth-mass planets.  For example, the Gliese 581 system hosts super-Earths.  One of these super-Earths exists on the outer edge of the HZ and orbits its host star in a tidally-locked configuration~\citep{2007A&A...476.1365V}.  Tidal locking occurs when one side of a planet constantly faces its host star. Around low mass stars, a habitable planet must have a small semi-major axis to exist in the HZ.  As a result of tidal friction, planets will experience tidal locking with their host stars on short timescales.  Consequently, planets will have one side constantly facing their host star, while the other side is in perpetual darkness.  Volatiles on the dark-side of the planet may freeze out.  Such an environment is thought to make a planet uninhabitable \citep{1964QB54.D63.......,1993Icar..101..108K}.  Habitability on planets that orbit low-mass stars has been evaluated to investigate the claim that they might be inhospitable to life.  Research shows that heat transfer mechanisms and certain atmospheric compositions could prevent volatiles from freezing out under particular conditions \citep{1997Icar..129..450J,2007AsBio...7...30T}.  Furthermore, tidal heating could permit plate tectonics that would allow for the recycling of CO$_2$ to prevent a runaway greenhouse effect \citep{2009ApJ...700L..30B}. Habitability on planets that are tidally locked with their host stars is likely to be different than habitability on planets that are not tidally-locked.  Therefore, tidally-locked and non-locked planets are investigated independently in our model.

\par
Each star in our model is assigned an appropriate stellar mass, which is used to scale the probability of having a habitable planet. With regards to tidal locking, the radius within which an Earth-like planet would become tidally locked is determined, using the same assumptions and consequent tidal-locking line as \citet{1993Icar..101..108K}.  The tidal-locking line describes the radius at which a planet will be locked into synchronous rotation by 4.5 Gyr for a particular stellar mass and orbital distance.  Given the orbital distance of a planet and its host star's mass, we estimate the probability of a planet becoming tidally-locked using data from the simulations of \citet{2005ApJ...626.1045I}.  In the respective bins in our model, all stars with $M_*$=0.08-0.5 host tidally-locked HZ planets; 43\% of stars with $M_*$=0.5-0.8 host tidally-locked HZ planets, and all stars with $M_*$=0.8-1.5 host non-locked HZ planets.  In other words, all M stars host tidally-locked planets, K stars host a mix of locked and non-locked planets and all G, and F stars host non-locked planets.

\par
Other factors exist that may reduce habitability around low-mass stars.  Planets are likely to receive X-ray and UV radiation fluxes at a much higher rate around low mass stars than around solar-mass stars of the same age \citep{2007AsBio...7..185L,2007A&A...476.1373S}.  In addition, planets that form in the habitable zones of low-mass stars may be deficient in volatiles as a result of high temperatures and collision rates during the planetary accretion phase \citep{2007ApJ...660L.149L}.  Furthermore, there is a tendency for low-mass stars to produce low-mass planets, and these planets are less likely to sustain substantial atmospheres and plate tectonic activity that is probably required for complex life \citep{2007ApJ...669..606R}.
\par
Other properties that could limit the lifespan of the biosphere are not modelled in the present study. For example, the HZ around a star changes as a function of time, but is not implemented here.  Furthermore, a significant loss of radiogenic heating within the interior of a planet could lower the CO$_2$ levels below the threshold required for photosynthesis \citep{2001cerc.book.....C}. Hence, the biosphere may not be conducive to complex life.

\subsection{Sufficient Time for the Evolution of Land-Based Complex Life}\label{life}
The only known examples of life in the universe reside on the Earth.  We do not know if the time required for the biological evolution of complex life on Earth is representative of the average time required on other planets.  To be able to quantify the number of habitable planets in the Galaxy, and compare our work to related studies, we assume that the biological evolution of life on other planets is identical to that of the Earth.  This assumption is highly speculative; however, we cannot extrapolate to other evolutionary paths without knowledge of complex life on other planets.  Our model assumes that complex life is animal life that dwells on land.  We assume that all of these planets have land; however, if water worlds prove to be common, we will have overestimated the number of habitable planets.
\par
The impact of a SN on life that inhabits water is not fully known; however, the probability of any catastrophe such as asteroid impacts, gamma ray bursts etc. entirely destroying all microbial life is very low.  On the other hand, a SN explosion in close proximity to the Earth would have the capability of sterilizing all land-based complex life.  This subsection describes how we model the timescales for the emergence of complex life, while accounting for the impacts of SN sterilizations.

\subsubsection{Constructing Earth-Analogs}\label{earth_analog}
\par
For complex life to exist on a habitable planet in our model, additional conditions must be met.  These conditions are based on the milestones and consequent timescales that occurred on Earth and are outlined in Table~\ref{ex:complex_life}.

\par
The buildup of oxygen in Earth's atmosphere coincided with the rise of multicellular life, and reached present day levels between 500 and 1000 Mya \citep{1998ASPC..148..449M}.  Fossil evidence of animal life occurs at $\sim$600 Mya, and molecular clock dates of the diversification of animals occurs at $\sim$1 Gya \citep{2005AsBio...5..415C}.  Given that the age of the Earth is $\sim$4.55 Gyr \citep{1995GeCoA..59.1445A}, we assume that the rise of animal life occurred $\sim$4 Gyr after the planet's formation.

\par
The destruction of ozone in the atmosphere of a planet caused by a nearby SN exposes land-based complex life to its host star.  The Earth's ozone layer was sufficiently developed $\sim$2.3 Gya \citep{2005AsBio...5..415C}; therefore, if an Earth-analog is bombarded by the flux from a SN before this time, the event has no effect on the development of complex life.  The emergence of animal life is assumed to be correlated with the rise of oxygen. Therefore, 1.55 Gyr of continuous ozone is required for the emergence of complex life.  If a SN occurs in the simulation between the formation of the ozone layer and the formation of animal life, we assume that the planet does not develop animal life until 1.55 Gyr of time elapses without interruption from ozone depletion events\footnote{For example: The ozone layer forms at 2.45 Gyr on each Earth-analog.  The Earth-analog is sterilized by a SN at T=2.7 Gyr.  The planet is not considered habitable until 4.25 Gyr in the simulation.}.  We require our model to reproduce major events that transpired on Earth to avoid speculation about life on extra-solar planets.
\par
As mentioned above, we ignore the production of stars in clusters.  SNII that would normally occur within a cluster and sterilize those young planets that have formed around stars in the cluster do not have a significant impact on habitability.  These SNII have a benign effect on habitability, as they occur well before ozone can form on our Earth-analogs.  Those stars and subsequent planets nearby a SNII that are not part of the cluster that contains the SNII are of a distribution of ages and therefore are expected to experience sterilizing events. Planets orbiting these stars may have had sufficient time for the buildup of ozone and rise of complex life.  We believe it is reasonable to assume that the background mean density of stars near the SNII that are not part of the cluster is a fair representation of stars we expect to be sterilized by the SN.  This is a realistic assumption as star clusters dissipate on short timescales, as their stars become well mixed within the disk.  We expect that an excess of stars above the mean stellar density at a given position as a result of clustering is superfluous, as this population of planets is not negatively affected by the SNII due to their young age.  Therefore, we believe that we capture the relevant SNe sterilizations despite the fact that we do not model clusters.

\subsubsection{Timescales for the Recurrence of Land-Based Complex Life}
\par
A planet is considered habitable if it satisfies the conditions necessary for complex life as described in \S~\ref{earth_analog}.  After these conditions are met, we monitor the planet to see if they are maintained.  If a sterilizing event occurs, we evaluate if land-based complex life can re-emerge.  For a planet to be become habitable a successive time, the reconstruction of the ozone layer and the evolution of animal life must occur.
\par
The rise in oxygen on the Earth occurred at $\sim$2.3 Gya, and evidence of cyanobacteria occurs at $\sim$2.7 Gya.  Therefore, the best-case scenario assigned for the reconstruction of the ozone layer is 0.4 Gyr, if cyanobacteria populations do not become extinct after a sterilizing event.  The worst-case scenario for the reconstruction of the ozone layer is assumed to be 2.25 Gyr, where the ozone layer builds up along the same timescale that it took from the formation of Earth (4.55 Gya) to the first wave of large-scale oxygenation at 2.3 Gya.  Therefore, the time period required for reconstituting ozone in the atmosphere is determined by choosing a value uniformly in the range [0.4,2.25] Gyr.

\par
The restoration of the ozone layer is assumed to be a prerequisite for land-based animal life.  We assume that the best-case scenario for the redevelopment of animal life is 0 Gyr, as complex life might survive in the oceans and immediately inhabit land when sufficient ozone is present.  The worst-case scenario is 1.55 Gyr, wherein a continuous ozone layer is required before the redevelopment of animals.  Therefore, the duration of time that elapses before animal life evolves and inhabits land is determined by choosing a value uniformly in the range [0,1.55] Gyr.

\par
It is not clear how long it would take for the reconstruction of the ozone layer and consequent colonization of land by complex life on exoplanets, and we can only make estimates of these values in the context of Earth.  The timescales for the recurrence of land-based complex life described here are therefore more speculative than other observable properties that are considered in the model.

\section{Results}\label{results}
\par
We have described a model of the Milky Way which reproduces well established physical properties of the Galaxy and have produced a set of criteria that allows us to predict where and when complex life might emerge.  The parameters and criteria are summarized in Table~\ref{ex:table_summary}. We now present the results from our simulations.
\par
The results are organized as follows. \S~\ref{results_sn} describes the influence of SN sterilizations on habitability in the Galaxy without considering planet formation. \S~\ref{results_complex_life} includes constraints on the time required for complex life to evolve, without considering planet formation. In \S~\ref{results_planets} we apply the SNe sterilizations and timescales for complex life to planet formation. \S~\ref{results_radial_dist} discusses the GHZ in terms of radial distance and height above the midplane. \S~\ref{kinematics} discusses the impact stellar kinematics would have on our models. In \S~\ref{IMF_habitability} we examine the impact of the IMF on habitability.  Finally, \S~\ref{results_compare} compares our results with relevant studies.

\subsection{The Effect of Supernovae Sterilizations on Habitability}\label{results_sn}
\par
In this section, stars that are close to SN events are investigated and planet formation is ignored.  We find that the majority of stars in our Galaxy will be bathed in flux by a nearby SN event during their lifetimes. Figure~\ref{fig:sterilization_radial} shows the number of stars in the Galaxy that remain unsterilized.  In this context, an unsterilized star refers to a star that has not been sufficiently close to a SN event in its lifetime.  Across all models, $\sim$27\%-36\% of all stars in the Galaxy remain unsterilized.  The lowest fraction of unsterilized stars in all models is located at R$\sim$2.5 kpc due to the high stellar density in the region. We predict that this fraction will be even lower at R$<$2.5 kpc, although we do not model this area.   To observe how sensitive stars are to sterilizations, we double the sterilization volume of SNe in the simulation.  When the sterilization volume is doubled, $\sim$23\% of all stars in the Galaxy remain unsterilized (Figure~\ref{fig:sterilization_radial}-Upper panel).  Doubling the sterilization volume does not double the number of stars that experience a sterilization during their lifetimes because some stars are sterilized multiple times.  This analysis indicates that a major increase in sterilization volume is unlikely to significantly reduce habitability within the Galaxy.

\par
The probability that a star remains unsterilized by a SN event ranges from $\sim$10-90\% in terms of galactocentric distance, and $\sim$10-95\% in terms of distance above the midplane (Figure~\ref{fig:fraction_of_non-sterilized_stars-radial}) across all models.  We can conclude from this finding that the fraction of planets expected to survive a SN event is much lower towards the centre of the Galaxy, and within and close to the midplane.  Comparing Models~1 and 4, which set the upper and lower limits of non-sterilized stars at R=8 kpc, we find that between $\sim$53\% and $\sim$36\% of all stars remain unsterilized at this radial position.
\par
SNII and SNIa are implemented separately to reflect the differences in their sterilization distances and progenitors. We find that in Model~1 SNII are responsible for more sterilizations than their SNIa counterparts.  There are $\sim$10.3$\times$ more SNII than SNIa; however, SNII only lead to $\sim$1.8$\times$ more sterilizations. We estimate that an individual SNIa is $\sim$5.7$\times$ more lethal than a SNII on average.  In Model~4, there are $\sim$11.7$\times$ more SNII than SNIa; however, SNII only lead to $\sim$2.1$\times$ more sterilizations.  An individual SNIa is roughly 5.6$\times$ more lethal than a SNII in Model~4.  Comparing the overall number of SNe, there are approximately 2$\times$ more SNII and  1.8$\times$ more SNIa in Model~4 than Model~1.  The Kroupa IMF in Model~4 produces a much higher number density of SNe, despite the lower number of stars in the model in comparison to Model~1.

\subsection{Timescales for Complex Life Around Main Sequence Stars}\label{results_complex_life}
\par
Disregarding planet formation, we add a 4 Gyr time constraint corresponding to the timescale required for the emergence of complex life.  In this subsection we analyze the positions of stars that are suitable for complex life when considering SNe sterilizations.  Hence, the star must have 1) a main sequence lifetime of at least 4 Gyr, and 2) remain unsterilized for a 4 Gyr period.  If a star is sterilized during this period, the biospheres of orbiting planets could be affected.  In this scenario, SNe will have the effect of delaying the emergence of complex life.  The 4 Gyr constraint on those stars suitable for habitable planets is relaxed when planet formation is considered in Section~\ref{results_planets}, where the assumptions for the development of complex life in Section~\ref{life} are adopted.
\par
Figure~\ref{fig:fraction_of_non-sterilized_stars_and_those_habitable_for_4_Gyr} presents the fraction of stars that are unsterilized for at least one period of 4 Gyr.  Interestingly, the inner Galaxy contains a fraction of sterilized stars that nevertheless exhibit a 4 Gyr unsterilized period that is suitable for complex life. Across all models, the lowest fraction of stars that are 4 Gyr of age and remain unsterilized are located in the inner Galaxy (Figure~\ref{fig:fraction_of_non-sterilized_stars_and_those_habitable_for_4_Gyr}-Upper panel).  This is expected, as the majority of stars in the inner Galaxy are sterilized (Figures~\ref{fig:sterilization_radial} and~\ref{fig:fraction_of_non-sterilized_stars-radial}).  The reverse phenomenon is demonstrated in the outskirts of the Galaxy in all models; the majority of stars suitable for complex life are unsterilized and at least 4 Gyr of age.  Due to the declining stellar density as a function of radial distance, there are more stars in total that would permit the emergence of complex life in the inner Galaxy than the outer Galaxy.  A high stellar density corresponding to a high SN rate would completely inhibit the emergence of complex life.  At R$\gtrsim$2.5 kpc there are no locations where this occurs, given the assumptions made in our model.
\par
In the middle and bottom panels of Figure~\ref{fig:fraction_of_non-sterilized_stars_and_those_habitable_for_4_Gyr}, the fraction of stars with 4 Gyr lifespans declines at R$\gtrsim$12 kpc, as many of the stars in the region are not 4 Gyr of age; therefore, these stars are less habitable than those at R$<$12kpc. Furthermore, the fraction of stars that have 4 Gyr periods where they are unsterilized by SNe across the disk is higher in Model~1 in comparison to Model~4, as the Salpeter IMF produces fewer SNe than the Kroupa IMF.

\subsection{The Effect of Metallicity on Planet Formation With Respect to Habitability}\label{results_planets}
\par
We now present our model of habitability with regards to planet formation as influenced by the metallicity gradient in the disk of the Milky Way galaxy.  Section~\ref{results_complex_life} discussed stars that are good candidates for habitable planets.  Two criteria in assessing the habitability of these stars were implemented: 1) the star must have a main sequence lifetime of at least 4 Gyr and 2) the star must remain free from sterilizing SNe events for a period of 4 Gyr.  Requiring criteria (2) is a strict prerequisite and does not reflect the impact SNe would have on our Earth-analogs given the major events that have transpired in Earth's history.  For example, a SN would have no effect on a planet that has not formed an ozone layer.  These milestones were reviewed in \S~\ref{life}, and the results of their implementation are presented in this section.  In this section, we highlight the results of Models~1 and 4 in parts of our analysis, as they most closely constrain the local number density in the solar neighbourhood.  Note that the Galaxy is best represented by the Kroupa IMF, and therefore we can expect Model~4 to best represent habitability within the Galaxy.

\par 
We find that the greatest number of habitable planets exist in the inner Galaxy in all of our models.  More specifically, 50\% of the habitable planets lie at R$<$4.1 kpc and R$<$4.4 kpc in Models~1 and 4 respectively.  The high stellar number density towards the centre of the Galaxy and the high degree of metallicity are responsible for permitting more stars to have habitable planets in this region in comparison to the middle and outer Galaxy (Figure~\ref{fig:Number_of_habitable_planets-total,tidal,non-tidal,function_of_radial}).  Moreover, star formation occurs earlier in the inner Galaxy, permitting longer periods for the emergence of complex life.  Furthermore, when comparing the age distribution of stars in the inner, middle, and outer regions of the Galaxy, the inner region allows for a greater chance of attaining habitable conditions in comparison to younger stars in the middle and outer regions of the disk.  The high stellar number density in the inner Galaxy and subsequent sterilization rate does not completely frustrate the emergence of complex life in the region.  As mentioned previously, only one major planet forms around a star in the work of \citet{2005ApJ...626.1045I}; therefore, the number of habitable planets is underestimated here.

\par
With respect to the fraction of habitable planets, we predict that between $\approx$1.2\% (Model~4) and $\approx$1.7\% (Model~1) of all stars host a habitable planet.  More specifically, in Model~1, 1.60\% of stars host a tidally-locked HZ planet and 0.09\% of stars host a non-locked HZ planet.  In Model~4, 0.9\% of planets are tidally locked, and 0.3\% are in a non-locked configuration with their host star.  The Salpeter IMF (Model~1) leads to more low mass stars whose companions become tidally locked on short timescales.  However, the Kroupa IMF (Model~4) leads to higher mass stars, permitting more planets in a non-locked configuration. In our model, the mix of locked and non-locked planets is constant throughout the Galactic disk.  The area with the greatest fraction of habitable planets over all epochs is located towards the centre of the Galaxy, as shown in Figure~\ref{fig:fraction_of_stars_with_habitable_planet-radial}.

\par 
From Sections~\ref{results_sn} and~\ref{results_complex_life}, we see that SN sterilizations on their own make the inner Galaxy the least hospitable for complex life.  However, regarding the planet-metallicity correlation without the effects of SNe makes the inner Galaxy the most hospitable for complex life.  When both factors are taken into account, the inner Galaxy is $\sim$10$\times$ more hospitable than the outer Galaxy (Figure~\ref{fig:fraction_of_stars_with_habitable_planet-radial}).  This finding indicates that the impact of metallicity on planet formation appears to dominate over the effects of SN sterilizations.  Furthermore, the inside-out scenario of Galaxy formation permits the inner region to be more habitable than the outskirts. Neither SN sterilizations nor metal-poor environments are capable of rendering any region inhospitable to complex life at the present day.

\par 
With respect to time, we investigate 1) the formation date (birth date) of habitable planets, and 2) the time in a planet's history when it becomes habitable.  Figure~\ref{fig:contour_HP_birth_date} shows the locations and birth dates of planets that are habitable at the present day.  All of our models indicate that most of the habitable planets are located in the inner Galaxy.  We find that our solar system is younger and distant from the densest regions that host habitable planets.  In addition, our results indicate that the average birth date of planets occurs 4.06 Gya, and 3.94 Gya in Model~1 and 4 respectively. Thus, many planets are too young to permit the evolution of complex life.  Over the next Gyr, many more planets are expected to attain habitable conditions given the assumptions made in our model.

\par
With respect to when planets attain habitable conditions, the top panel of Figures~\ref{fig:Model1-habitability_history} and \ref{fig:Model4-habitability_history} demonstrate that habitable planets have emerged for the past $\sim$6 Gyr near the Galactic centre (R$\gtrsim$2.5 kpc), whereas they have become habitable at $\sim$4 Gya in the middle region, and at $\sim$1-2 Gya in the periphery.  The SFR experienced in the last few billion years, coupled with increasing levels of metallicity, suggests that many more planets will be conducive to complex life in the future.

\par
Tracing the habitability history of planets over all epochs shows that at the present day, the fraction of stars hosting planets with habitable conditions is greatest at R$\sim$2.5 kpc (Figures~\ref{fig:Model1-habitability_history} and \ref{fig:Model4-habitability_history}- Top Right panel).  The inner Galaxy has the greatest fraction of stars with a habitable planet integrated over all periods of time (Figure~\ref{fig:fraction_of_stars_with_habitable_planet-radial}), and at the present day.


\subsection{The Location of the GHZ}\label{results_radial_dist}
\par
The GHZ is thought to be affected by an inner boundary that is determined by hazards to planetary biospheres, and an outer boundary set by the minimum amount of metallicity required for planet formation \citep{2001Icar..152..185G}. This outlook describes the GHZ as a function of radial distance.  However, we model concentrations of habitable planets with respect to galactocentric distance ($R$) and height above the midplane ($z$).
\par
The bottom left panels of Figures~\ref{fig:Model1-habitability_history} and \ref{fig:Model4-habitability_history} illustrate that the greatest number of habitable planets over all epochs is located in the inner Galaxy, within and surrounding the midplane.  This region coincides with the greatest stellar number density in the model.  Most planets in this densely populated area will be sufficiently close to SNe, but the frequency of ozone depletion events is not high enough to permit large volumes of the Galaxy to be sterile for long timescales.
\par
In the bottom right panels of Figures~\ref{fig:Model1-habitability_history} and \ref{fig:Model4-habitability_history} it is observed that the region with the greatest fraction of stars with habitable planets over all epochs is located in the inner Galaxy, well above the midplane, centred at a height of z$\sim$1.5 kpc.  The same radial mix of metals at this location above the midplane combined with the low stellar density in comparison to the density found closer to the midplane suggests that SNe have a significant impact on habitability.  The fraction of stars with habitable planets above the midplane is a factor of a few greater than the fraction at the midplane at R$\sim$2.5 kpc.
\par
We observe that the greatest number density and fraction of habitable planets is located in the inner Galaxy, integrated over all epochs.  These findings suggest that our location in the Galaxy is not particularly favourable under this GHZ paradigm.  We find that the GHZ, defined as the position with the greatest number of complex life supporting habitable planets, is located in the inner Galaxy, within and surrounding the midplane.

\subsection{Stellar Kinematics}\label{kinematics}
\par
The motions of stars are not considered in our model. Solar systems passing through dense regions of the Galaxy are more likely to be sterilized by SNe.  The lack of stellar kinematics would be a major concern if we found a continuously sterilized zone (CSZ), wherein an entire section of the Galaxy is uninhabitable as a result of transient radiation events, effectively sterilizing stars that pass through the region. However, the fraction of habitable planets in the densest region (at the midplane in the inner Galaxy) is not significantly lower than most of the other parts of the Galaxy. Therefore, if the orbits of stars were modelled in the present study, we would not expect a major decrease in habitable systems as a result of traversal through high density regions.  Given the results presented in this study, neither radial or vertical motion presents a major concern.  In the radial case, if a star in the midplane (z$\sim$0 pc) moves inwards towards the Galactic centre, the star will be in an area where the fraction of stars with habitable planets increases (Figures~\ref{fig:Model1-habitability_history} and \ref{fig:Model4-habitability_history}-Bottom left panels), therefore, the SNe rate cannot be great enough to inhibit habitability in the majority of solar systems with radial components to their motion.  In the vertical case, if a star in a low density region above the midplane (e.g. R$\sim$2.5 kpc and z$\sim$2 kpc) plunges towards the midplane, the fraction of habitable planets only decreases by a factor of $\sim$2-3 (Figures~\ref{fig:Model1-habitability_history} and \ref{fig:Model4-habitability_history}- Bottom right panels).  The slightly decreasing fraction of habitable planets towards the midplane at R$\sim$2.5 kpc suggests that vertical motion cannot be responsible for a major decrease in habitability.  If vertical stellar motions are taken into account, the vertical gradient of the fraction of habitable planets is likely to be reduced.  Further research on habitability in the Galactic bulge in combination with stellar kinematics is necessary to determine if dense regions have a significant impact on habitability. If a large CSZ exists in the inner disk or the bulge, stars that reside in the region, or those stars in eccentric orbits that pass through the region, may not have planets that can support complex life as defined in this study.

\subsection{The Impact of the IMF on Habitability}\label{IMF_habitability}
\par
We present four models to see how sensitive our model is to varying properties of the Galaxy. A Salpeter IMF is utilized to assign masses to the stars in Models~1 and 3 and the Kroupa IMF is used in Models~2 and 4.
\par
Comparing Models~1 and 4, we see that the former has a lower SNe rate than the latter.  There are $\sim$1.8$\times$ more SNII and $\sim$1.7$\times$ more SNIa in the models with the Kroupa IMF in comparison to those with the Salpeter IMF.  Taking the SNe rates in consideration on their own makes the Galaxy less hospitable for complex life when the Kroupa IMF is used to assign masses to our stars.  However, when comparing the fraction of stars that host habitable planets in a non-locked configuration (Figure~\ref{fig:fraction_of_stars_with_habitable_planet-radial}), the Kroupa IMF produces more high mass stars permitting many more planets to orbit their hosts in a non-tidally locked configuration in their HZ in comparison to planets formed in the models containing the Salpeter IMF.  Evidence suggests that the Galaxy follows the Kroupa IMF; therefore, we expect $\sim$0.3\% of stars to host a planet capable of harbouring complex life that orbits in a non-locked configuration.


\subsection{Comparison with other Studies}\label{results_compare}
\citet{2008SSRv..135..313P} suggests that the concept of the GHZ may have very little significance, and that it should be considered only as a framework to organize our ideas about life in our Galaxy.  While it is understood that further research is required to better answer questions related to habitability on the Galactic scale, we do believe that these factors can be quantified at present with useful results.  It is certain that a sufficiently nearby SN would have a sterilizing impact on land-based life on Earth, and the evolution of biologically complex life took a substantial amount of time on our planet.  Studies of this nature imply uncertainties that require us to make reasoned assumptions concerning habitability.  In the future, these assumptions will be replaced with observational fact.
\par
In this subsection we compare key differences between the present study and the related literature.  Unlike previous approaches, our simulation models each star individually. While such a star-based model incurs increased memory and computational costs, it potentially affords more realistic and informative results than traditional probabilistic models that aggregate over stellar populations.  The contrasting methods and results are reviewed, some of which have been described or alluded to in previous sections.

\subsubsection{Model Milky Way Galaxies}
\par
\begin{figure}[htp]
 \begin{center}
  \includegraphics[width=0.47\textwidth]{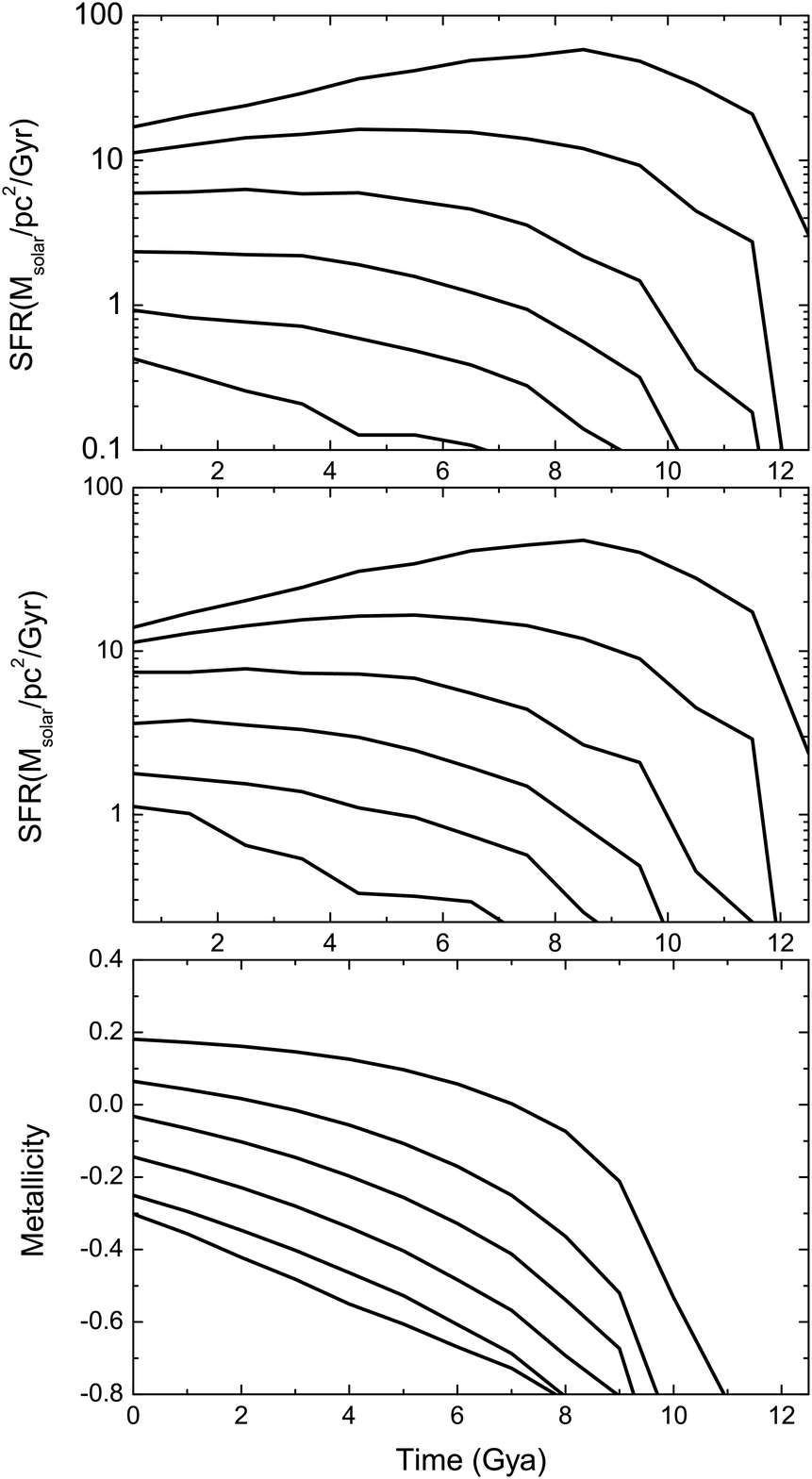}
 \end{center}
 \caption{Temporal profile of the SFR and metallicity in our model at six galactocentric regions: 2.5, 5, 7.5, 10, 12.5, 15 kpc ordered from top to bottom.  Upper panel: Temporal profile of the SFR in Model~1.  The curves indicate a burst of star formation in the inner Galaxy that declines with time.  The SFR at 7.5 kpc had a burst of star formation early in the history of the Galaxy and flattens within the last few Gyr.   Middle panel: Temporal profile of the SFR in Model~4.  The SFR is similar to that of Model~1, except the SFR is slightly lower in the inner Galaxy, and higher in the outskirts.  Lower panel: Temporal profile of the average metallicity in our models.  The inner Galaxy produces metals early as a result of the burst of star formation in the region, whereas the build up of metals takes longer in the outer Galaxy.}
  \label{fig:SFR-time}
\end{figure}
\begin{figure}[h]
  \begin{center}
   \includegraphics[width=0.47\textwidth]{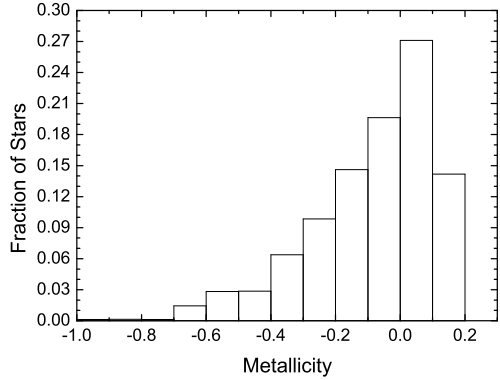}
  \end{center}
  \caption{The present day distribution of metallicity for all stars in our model that have formed at the solar radius (R$\sim$8 kpc).  This distribution is consistent with the model of \citet{2006MNRAS.366..899N}.}
  \label{fig:metallicity_dist_8kpc}
\end{figure}
\begin{figure}[h]
\center
\includegraphics[width=0.47\textwidth]{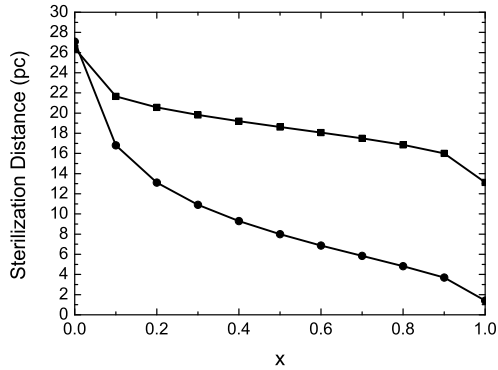}
\caption{The distribution of sterilization distances for SNII (circle) and SNIa (square).   The curves represent the distribution of sterilization distances, corresponding to the distribution of absolute magnitudes of SNII and SNIa.  The $x$ value is used to select a sterilization distance for a given SNII or SNIa, where $x$ is a random number generated between 0 and 1 in the Monte-Carlo simulation.  It is not surprising that the range of sterilization distances for SNIa is less than that of SNII, reflecting the view that SNIa are considered to be standard candles.  SNIa are more luminous on average and therefore have a greater average sterilization distance than SNII.}\label{fig:sterilization_distances}
\end{figure}
\begin{figure}[h]
\center
\includegraphics[width=0.47\textwidth]{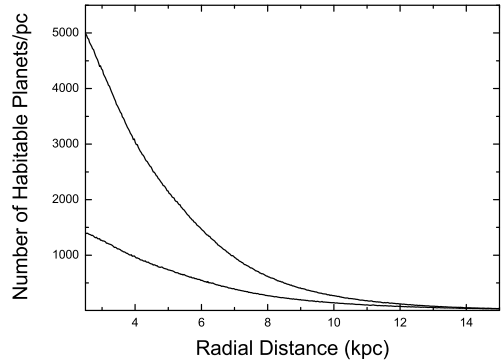}
\caption{The positions of all of the habitable planets formed over all epochs in Model~1 (upper curve) and Model~4 (lower curve).  These planets have $0.1~M_\oplus<M_p<10~M_\oplus$, exist in the HZ of their host stars and have survived the negative effects posed by Hot Jupiters.  The discrepancy between the number of planets formed in Models~1 and 4 is a function of the IMF, where the Salpeter IMF in Model~1 produces more stars, and hence, more planets than the Kroupa IMF utilized in Model~4.  This may give us a rough idea of the distribution of planets capable of hosting microbial life.}\label{fig:all_planets}
\end{figure}
\begin{figure}[!tbph]
  \begin{center}
    \includegraphics[width=0.55\textwidth]{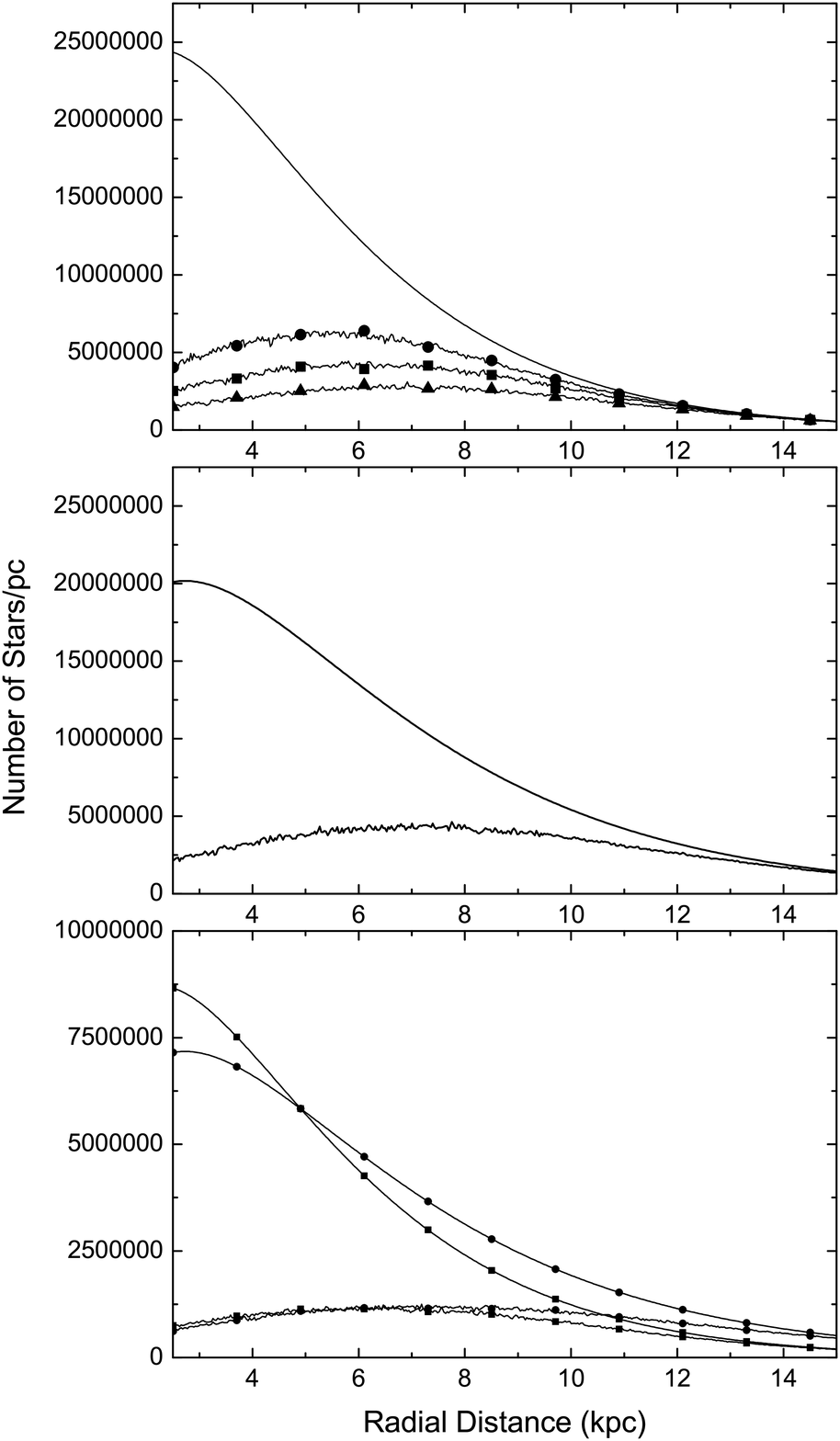}
  \end{center}
  \caption{The number of stars in the Galaxy that would not have their respective solar systems sterilized by a SN is plotted as a function of galactocentric distance for each model. Upper panel: Models 1, 1a and 1b are plotted.  The top curve represents the total stars in the model.  The curve with the square markers represents those non-sterilized solar systems with our default sterilization distance.  We find that $\sim$33.3\% of stars are never sterilized by a SN event.  Stars in the inner Galaxy are more likely to be sterilized than those in the outskirts.  Our sensitivity analysis indicates that when the sterilization volume is halved, $\sim$45\% of all stars in the Galaxy remain unsterilized (circle markers-Model~1a).  When we double the sterilization volume (triangle markers-Model 1b), $\sim$23\% of all stars in the Galaxy remain unsterilized.  Middle panel: Model~3 is plotted.  The top curve represents the total stars in the model.  In this model $\sim$36\% of all stars remain unsterilized (bottom curve).  Bottom panel: The total stars and unsterilized stars for Model~2 (square markers) and Model~4 (circle markers) are plotted. In Model 2, $\sim$27\% of all stars remain unsterilized, whereas in Model 4, $\sim$29\% of all stars remain unsterilized.}
  \label{fig:sterilization_radial}
\end{figure}
\begin{figure}[!tbph]
  \begin{center}
    \includegraphics[width=0.47\textwidth]{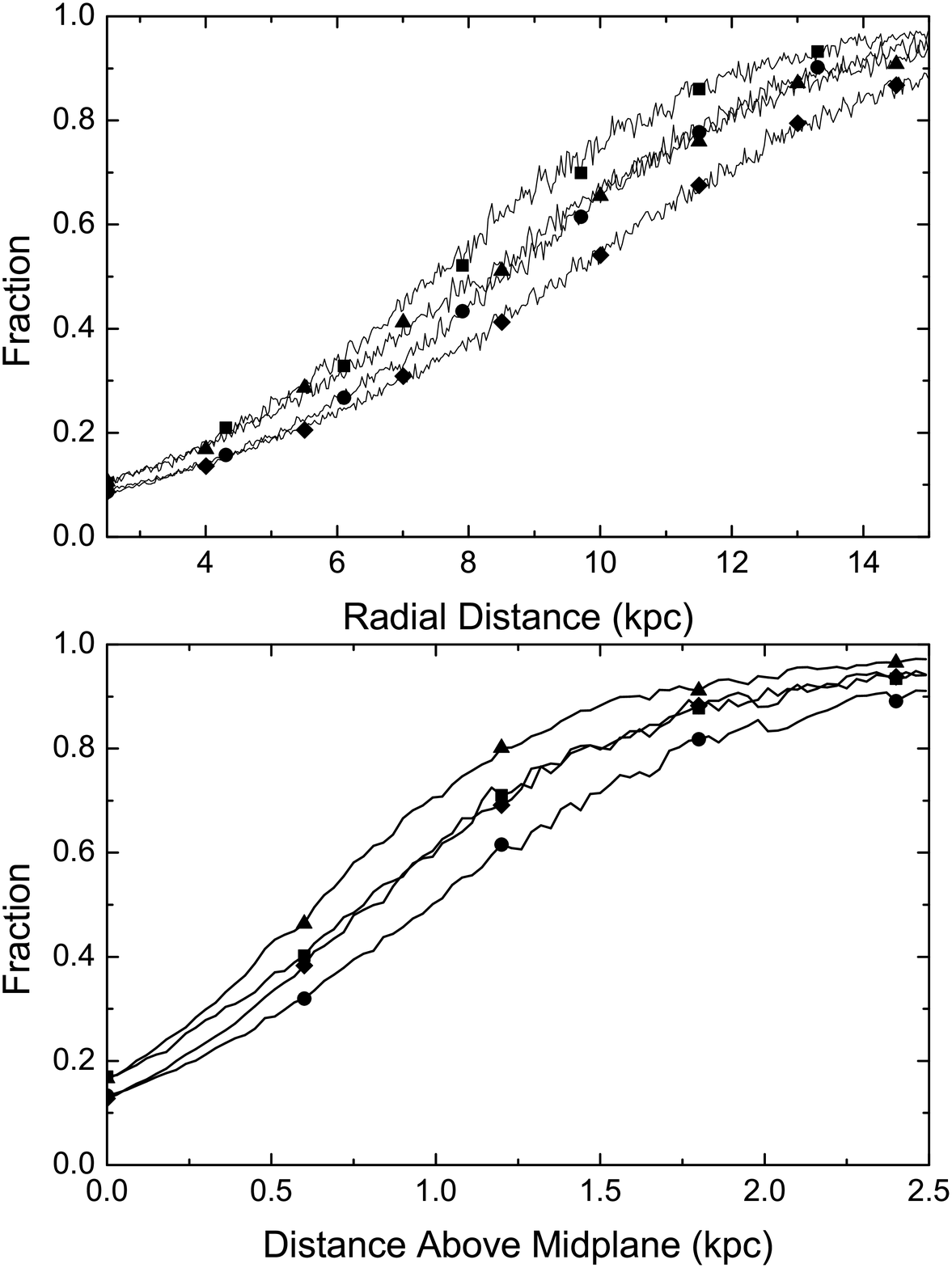}
  \end{center}
    \caption{Upper panel: The fraction of stars that are not sterilized by a SN event as a function of galactocentric distance.  Models 1-4 are labeled with the square, circle, triangle, and diamond markers respectively.  Only a small fraction of stars remain unsterilized by a SN event in the inner Galaxy, whereas almost all stars are unaffected in the outer Galaxy.  All models converge in the inner Galaxy where $\sim$10\% of all stars are not sterilized.  Likewise, in the outer Galaxy, all models converge, yielding $\gtrsim$90\% of all stars to remain unsterilized.  At R=8 kpc, between $\sim$36\% (Model~4) and $\sim$53\% (Model~1) of all stars remain unsterilized at this radius.  Lower panel: The fraction of stars that are not sterilized by a SN event as a function of height above the Galactic midplane.  Only a small fraction of stars remain unsterilized by a SN event at the midplane, whereas almost all stars remain unsterilized at z$\sim$2.5 kpc.}
  \label{fig:fraction_of_non-sterilized_stars-radial}
\end{figure}
\begin{figure}[htp]
  \begin{center}
    \includegraphics[width=0.47\textwidth]{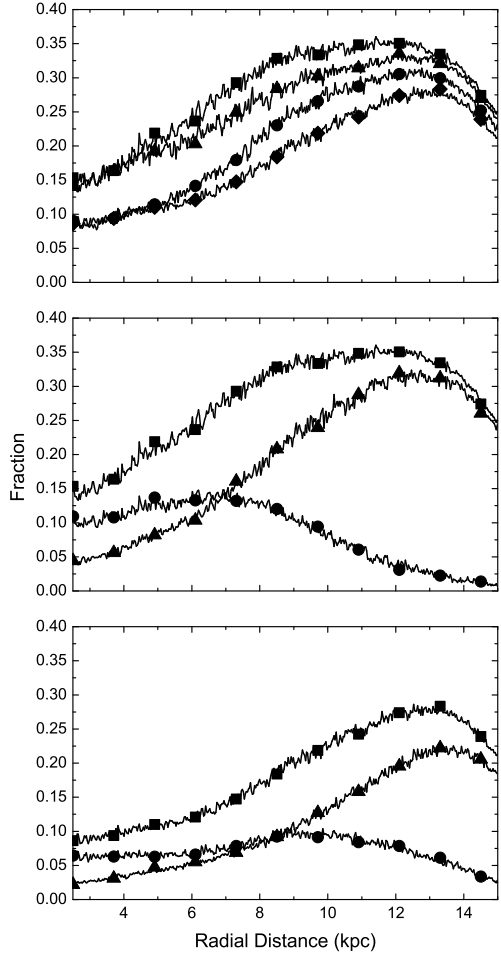}
  \end{center}
  \caption{Upper panel: The total fraction of stars that have 4 Gyr periods where they are unsterilized by SNe is plotted versus radial distance.  The square, circle, triangle, and diamond markers correspond to Models 1, 2, 3, and 4 respectively.  The inner Galaxy has the lowest fraction of stars suitable for complex life. At R=8 kpc, between $\sim$15\% (Model~4) and $\sim$30\% (Model~1) of stars have 4 Gyr periods where they are not sterilized by SNe.  Middle panel: Model 1 is plotted.  The curve with triangle markers represents the fraction of stars that are never sterilized; the curve with circle markers represents the fraction of stars that have been sterilized, but still have a period of 4 Gyr absent from sterilizing events; and the curve with square markers (triangle + circle) represents the total fraction of habitable stars, which have 4 Gyr periods unsterilized by SNe and is also shown in the upper panel.  The majority of stars suitable for complex life in the inner Galaxy have been sterilized.  Bottom panel: The same as the middle panel; however, Model~4 is plotted.  We find that roughly half of the stars that are capable of supporting complex life are never sterilized at 8 kpc.}
  \label{fig:fraction_of_non-sterilized_stars_and_those_habitable_for_4_Gyr}
\end{figure}

The buildup of the Galactic stellar mass reported in \citet{2008SSRv..135..313P} and the present study is roughly consistent.  The SFH employed in \citet{2008SSRv..135..313P}, \citet{lineweaver-GHZ}, and our model, has an early burst of star formation as illustrated in the upper and middle panels of Figure~\ref{fig:SFR-time}.  The average metallicity assigned to each star in our model (Figure~\ref{fig:SFR-time} - lower panel), is roughly consistent with the metallicity found in \citet{lineweaver-GHZ}.  The measure of metallicity in \citet{2008SSRv..135..313P} is different than ours; therefore, a direct comparison is not possible.

\subsubsection{Supernovae Rates and Sterilization Distances}
\par
Our research addresses the danger effect of SNe using a different approach than that of \citet{2008SSRv..135..313P} and \citet{lineweaver-GHZ}.  Numerous studies in the field show that SNIa and SNII have different distributions of luminosities \citep{2002AJ....123..745R,2006ApJ...645..488W}.  Our model accounts for this using a distribution of sterilization distances for SNII and SNIa.  Conversely, \citet{2008SSRv..135..313P} and \citet{lineweaver-GHZ} assess the SN danger factor as a time integrated supernova rate.  Treating SN events in our model on an individual basis within the pre-existing stellar population improves on the statistical approaches of other studies, as it is a more realistic, less Earth-centric method.  Our model does not assess the danger posed by SNe as normalized to the Earth's radial position.

\subsubsection{Metallicity and Planet Formation}
\par
There are few studies that predict the number of habitable planets in the Milky Way \citep{2001Icar..151..307L,2007AsBio...7..745B,2009Ap&SS.tmp..118G}.  We use a metallicity profile \citep{2006MNRAS.366..899N} to assign a metallicity to each star in the model.  We use this metallicity to model planet formation.  Considering that \citet{2008SSRv..135..313P} uses a constant probability of stars forming an Earth-like planet (40\%), as influenced by \citet{2001Icar..151..307L}, our model contains far fewer habitable planets.  Moreover, the danger of a Hot Jupiter inhibiting an Earth-like planet in our study is similar to \citet{2008SSRv..135..313P}, as we share the results by \citet{2005ApJ...622.1102F}, producing a much lower Hot Jupiter danger effect than that of \citet{lineweaver-GHZ}.

\subsubsection{Comparison of Results}
\par
The GHZ is an area in the Galaxy that contains stars with the highest potential to harbour complex life. \citet{lineweaver-GHZ} find the GHZ to be an annular region between 7 and 9 kpc at the present day.  Our results fundamentally disagree with \citet{lineweaver-GHZ}, as we find that the greatest number of habitable planets exist in the inner Galaxy.  However, we recognize that the model Galaxy produced by \citet{2003PASA...20..189F} and employed in \citet{lineweaver-GHZ} is more advanced than our model in certain regards.  Their model contains a more accurate depiction of the observed properties of the Galaxy such as considering the SFR in conjunction with spiral arm motions.
\par
We expect that as the metallicity increases, the entire disk is expected to harbor a greater number and fraction of habitable planets.  This result is similar to that of \citet{2008SSRv..135..313P}.  Moreover, we find that the number of habitable planets varies as a function of height above the midplane.  This was not investigated in \citet{lineweaver-GHZ} or \citet{2008SSRv..135..313P}, and therefore we observe that the morphology of the GHZ is not an annular region as suggested in \citet{lineweaver-GHZ}, or the entire disk as discussed in \citet{2008SSRv..135..313P}, but rather consists of a region highly dependant on radial distance, located near the centre of the Galaxy.
\par
Habitability may be reduced at R$\sim$2.5 kpc as a result of the high stellar density in combination with stellar motions, and will be investigated in future research concerning the Galactic bulge.  Nonetheless, our prediction that many habitable planets exist at the midplane in the inner Galaxy suggests that habitable conditions are possible in high density regions.

\par
The dominant paradigm of habitability on the Galactic scale is certain to change in this field as we learn more about our Galaxy and habitable planets.  Studies relating to the distribution of habitable planets, extraterrestrial life and colonization of the Galaxy \citep[see][]{2007AsBio...7..745B,2009IJAsB...8..121F} currently rely on the canonical model by \citet{lineweaver-GHZ}.  Our revised conception of the GHZ has implications for these and other related studies.

\section{Conclusions}\label{conclusion}
\par
We present a model of habitability within the Milky Way to predict the region(s) in the Galaxy that are expected to favour the emergence of complex life.  We do not find that the inner Galaxy is entirely inhospitable to life; in fact, the greatest number of habitable planets are found in this region.  The metallicity in the inner Galaxy produces a high planet formation rate for long timescales that dominates the negative impact SNe have on habitability.  Furthermore, we observe that over all epochs in our favoured Model (Model~4), 1.2\% of all stars in the Galaxy host a habitable planet (including both tidally locked and non-locked configurations).  The fraction of stars with a habitable planet ranges from $\sim$0.25\% in the outer Galaxy to $\sim$2.7\% in the inner Galaxy.  Considering that we find the greatest number of habitable planets to be located in and around the midplane at R$\sim$2.5 kpc, and that the greatest fraction of stars that support habitable planets exist above and below the midplane at this radial position, we find that the inner region of the Galaxy may support the greatest number of planets conducive to complex life.  More specifically, our findings are at odds with the notion that the GHZ is shaped like an annular ring at R$\sim$7-9 kpc at the present day, as found by \citet{lineweaver-GHZ}.  Similarly to~\citet{2008SSRv..135..313P}, we find the greatest number of habitable planets to exist in the inner Galaxy, with the exception that we observe habitable planets at this radial position to be strongly dependent on height above and below the Galactic midplane.
\par
Given that we find the greatest concentration of habitable planets to be located in the inner Galaxy, further research on the disk and overlapping bulge at R$\lesssim$2.5 kpc is warranted.  The inner boundary of the GHZ is defined by hazards to a planet's biosphere \citep{2001Icar..152..185G}; however, we find that an inner boundary does not exist when modelling the Galactic disk at R$\gtrsim$2.5 kpc.  By modelling the region at R$\lesssim$2.5 kpc, we might find an inner boundary, or we might not find a boundary at all.
\par
\par
\par
Studies of habitability on the Galactic scale will improve in the future as Earth-like planets are detected from studies such as the Kepler mission \citep{2008IAUS..249...17B}, and will benefit from an improved understanding of our Galaxy from the GAIA mission \citep{2001A&A...369..339P}, and others.
\par
The results of planet finding missions will yield estimates of the total number of planets in the HZ of their host stars across the Galactic disk.  The total number of habitable planets estimated here is not the focus of our research; rather, we highlight and predict that these planets exist preferentially in the inner Galaxy, and that they have the ability to survive SN sterilization events for periods conducive to the rise of biologically complex life.

\section*{Acknowledgments}
This material is based upon work supported by the National Aeronautics
and Space Administration through the NASA Astrobiology Institute under
Cooperative Agreement No. NNA08DA77A issued through the Office of
Space Science. DRP gratefully acknowledges the financial support of a Discovery Grant
from the Natural Sciences and Engineering Research Council (NSERC) of Canada.
This paper also benefitted from the insightful comments of Jon Willis.

\bibliographystyle{apj}

\newpage

\begin{figure}[p]
  \begin{center}
    \includegraphics[width=0.47\textwidth]{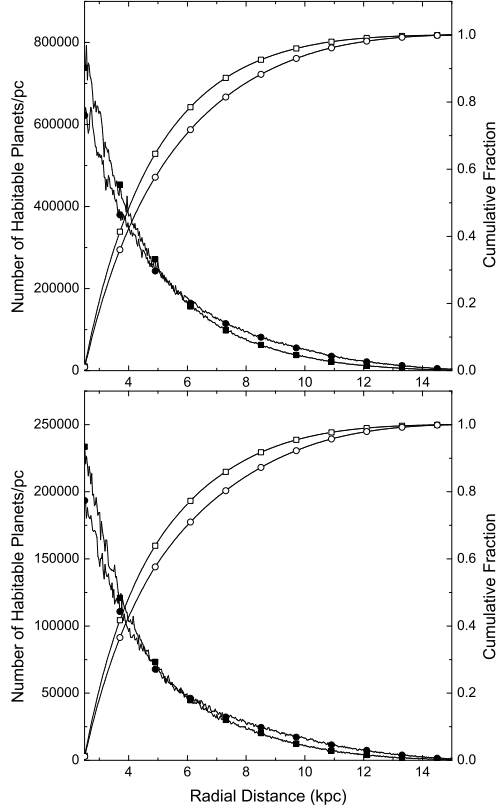}
  \end{center}
  \caption{The number of habitable planets is plotted versus radial distance.  Upper panel: Models~1 (square markers) and 3 (circle markers) are plotted (models with a Salpeter IMF).  The curves indicate the total number of habitable planets that occur over all epochs of time, including both tidally-locked and non-locked configurations.  The greatest number of habitable planets is located in the inner Galaxy.  In Model~1 $\approx$1.7\% of all stars host a habitable planet, 1.60\% of which are considered to be tidally locked to their respective host star and 0.09\% are in a non-tidally locked configuration.  We find that 50\% of the habitable planets in Model~1 lie at R$<$4.1 kpc.  Model~3 is similar to that of Model~1.  Lower panel: Models~2 (square markers) and 4 (circle markers) are plotted (models with a Kroupa IMF).  The results are very similar to those of Models~1 and 3, where the greatest number of habitable planets is located in the inner Galaxy. In Model~4 we find that $\approx$1.2\% of all stars host a habitable planet (0.9\% are tidally locked, and 0.3\% are in a non-locked configuration with their host star).
  }
  \label{fig:Number_of_habitable_planets-total,tidal,non-tidal,function_of_radial}
\end{figure}

\begin{figure}[p]
  \begin{center}
   \includegraphics[width=0.47\textwidth]{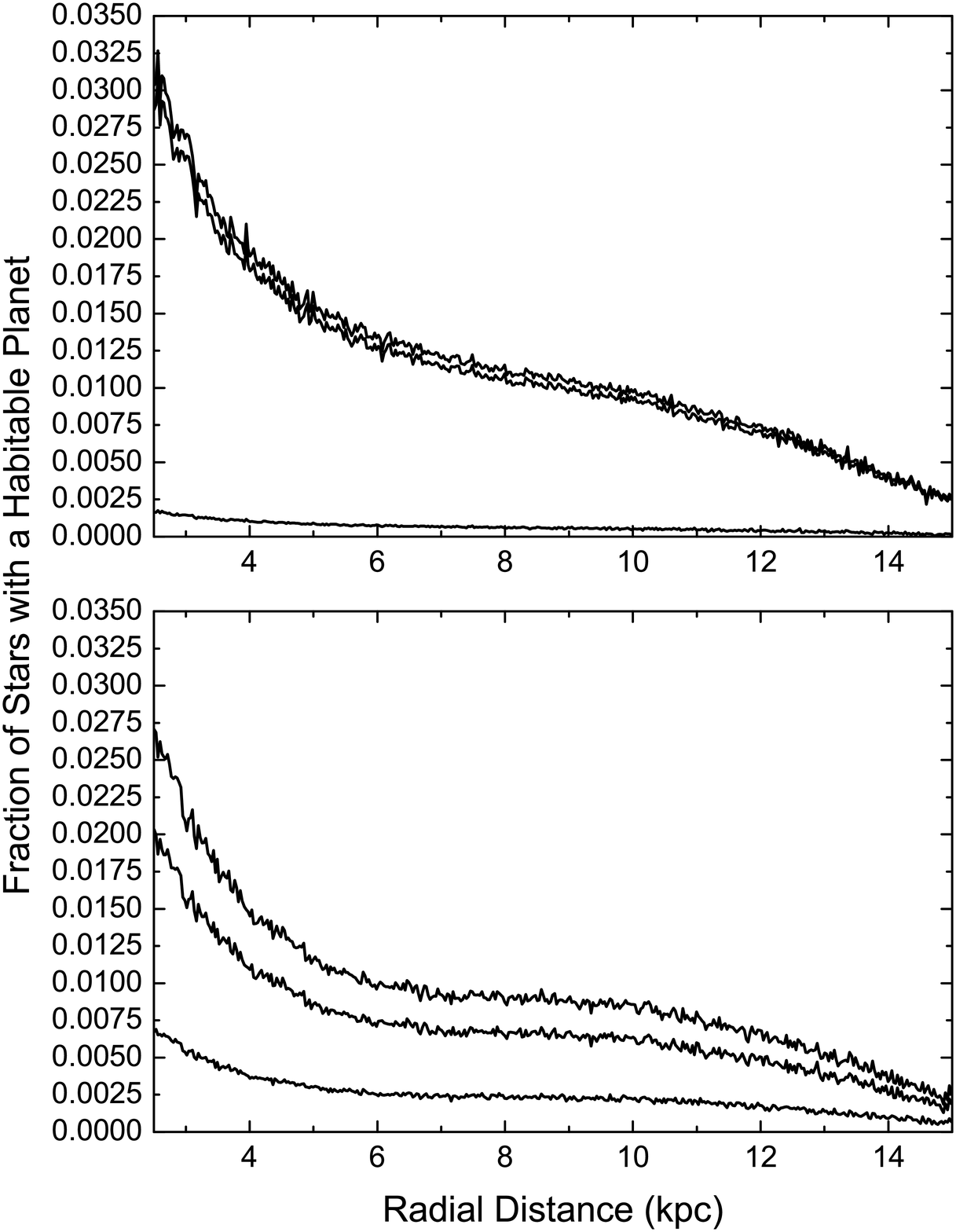}
  \end{center}
  \caption{The fraction of stars with a habitable planet over all epochs in the Galaxy is plotted as a function of radial distance.  Model~1 is plotted in the upper panel, and Model~4 in the lower panel.  The curves from top to bottom in each panel indicate the total number of habitable planets, those in a tidally-locked configuration, and those not tidally-locked to their host star.  There is a higher probability of a star having a habitable planet in the inner region of the disk.  The probability declines in the outer Galaxy, as the metallicity is not sufficient to produce high planet formation rates, despite the much lower SN sterilization rate.  Furthermore, planets form at an earlier time in the inner Galaxy than in the outskirts, allowing for more planets to attain habitable conditions, despite the higher SNe rate.  The distribution of stellar masses and subsequent probabilities of planet formation around stars of different masses indicates that the majority of habitable planets will be in a tidally-locked configuration.  From the plots, the IMF has a major influence on the fraction of planets that are not tidally locked to their host star.
  }
  \label{fig:fraction_of_stars_with_habitable_planet-radial}
\end{figure}

\begin{figure}[p]
  \begin{center}
   \includegraphics[width=1.1\textwidth]{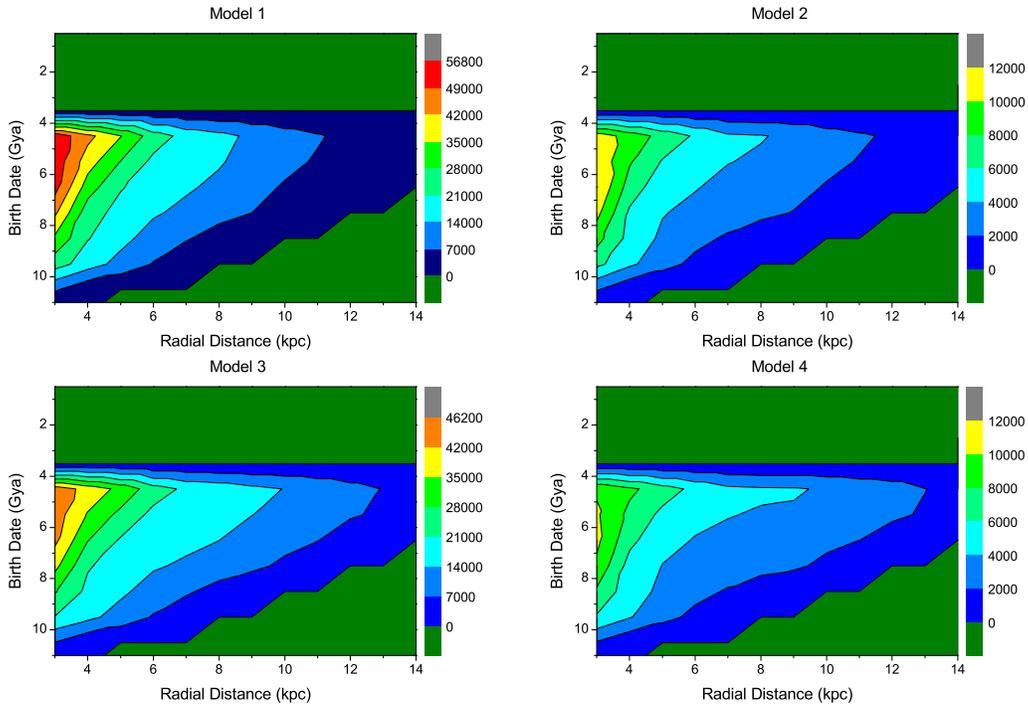}
  \end{center}
  \caption{The number of habitable planets per pc is plotted as a function of radial distance and birth date for each model.  We trace the history of each habitable planet to determine in which periods they remain habitable, and plot those that are habitable at the present day.  It is clear, given the assumptions made in our models, that habitable planets are most prevalent in the inner Galaxy.  Therefore, the inner Galaxy is expected to favour the emergence of complex life and the region is expanding outwards with time.  The majority of stars that host habitable planets in our model are expected to be older than the Sun.  This figure can be directly compared with Fig. 3 in~\citet{lineweaver-GHZ}.}
  \label{fig:contour_HP_birth_date}
\end{figure}

\begin{figure}[p]
  \begin{center}
 \includegraphics[width=1\textwidth]{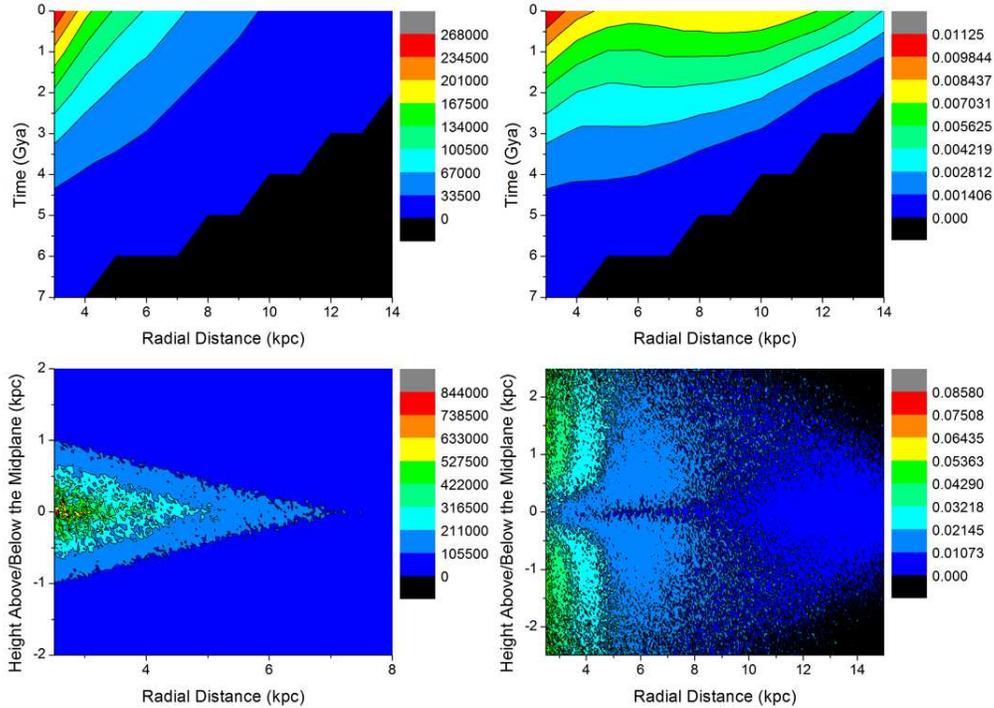}
  \end{center}
  \caption{In this Figure, Model~1 is plotted.  Top Left panel: The number of planets that are habitable (tidally-locked and non-locked) as a function of radial distance and time.  We trace the history of each habitable planet to determine in which periods they remain habitable.  It is clear, given the assumptions made in our model, that the number of habitable planets is greatest in the inner Galaxy, at all epochs.  Top right panel: The fraction of stars with a habitable planet as a function of time and radial distance.  We trace the history of each habitable planet to determine in which periods they remain habitable.  Given the assumptions made in the model, the inner region of the Galaxy exhibits the highest probability of having habitable planets at the present day.  At R$\sim$5-11 kpc, at the present day, the entire range has roughly the same probability of having habitable planets.  Lower Left panel: The number of habitable planets integrated over all epochs as a function of radial distance and height above the midplane. We predict that the position in the Galaxy with the greatest number of habitable planets is located in and around the midplane in the inner Galaxy.  Lower Right panel: The fraction of habitable planets integrated over all epochs as a function of radial distance and height above the midplane.  The region with the greatest fraction of habitable planets exists above the midplane in the inner Galaxy.  The high metallicity that produces a high planet formation rate and the lower stellar density that exists well above the midplane at this radial position permits a greater fraction of habitable planets to form.}
  \label{fig:Model1-habitability_history}
\end{figure}

\begin{figure}[tbp]
\begin{center}
\includegraphics[width=1\textwidth]{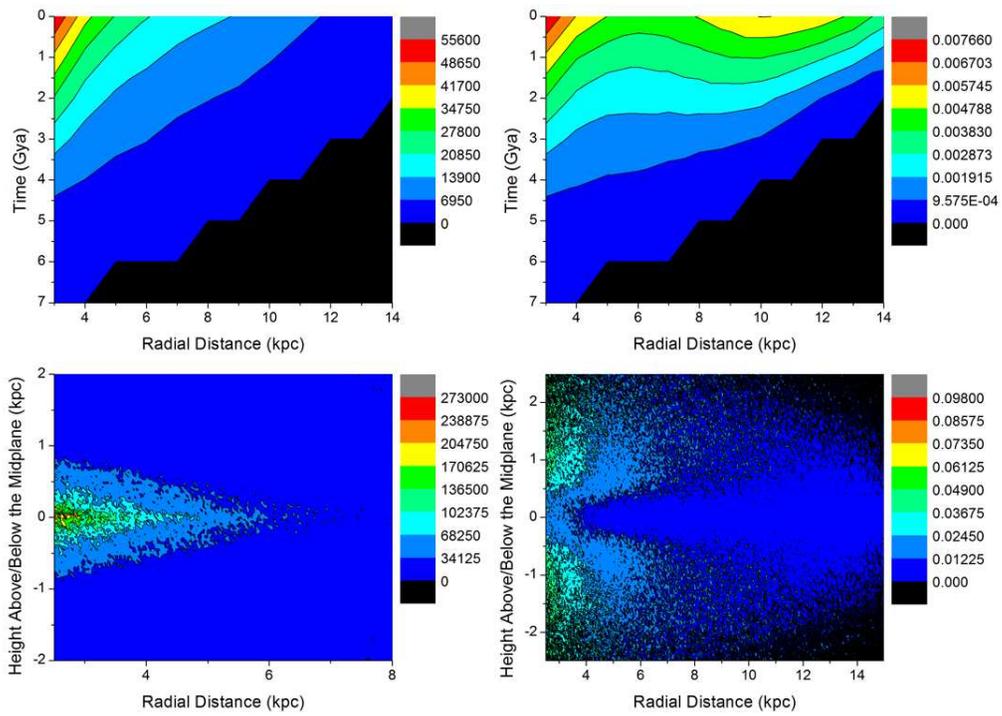}
\end{center}
\caption{The same as Figure~\ref{fig:Model1-habitability_history}, except Model~4 is shown.  The trends are very similar in both models.}
\label{fig:Model4-habitability_history}
\end{figure}

\pagestyle{plain}

\clearpage

\begin{centering}
\begin{table}
\begin{threeparttable}[b]
\caption{Description of the models}\label{ex:model_description}

\centering\small
\begin{tabular*}{1.25\textwidth}{cccccc}
\hline\hline
Model\tnote{1}&
Stellar Number Density Distribution&
IMF&
Sterilization Volume&
Number of Stars\tnote{2}&
n(30 pc, 8 kpc)\tnote{3}\\
\hline
1&\citet{2006ima..book.....C}&Salpeter&1&1.03&0.155\\
\hline
2&\citet{2006ima..book.....C}&Kroupa&1&0.366&0.055\\
\hline
3&\citet{2008ApJ...673..864J}&Salpeter&1&1.13&0.220\\
\hline
4&\citet{2008ApJ...673..864J}&Kroupa&1&0.402&0.078\\
\hline
1a&\citet{2006ima..book.....C}&Salpeter&0.5&1.03&0.155\\
\hline
1b&\citet{2006ima..book.....C}&Salpeter&2&1.03&0.155\\
\hline

\end{tabular*}

\begin{tablenotes}
\item [1] All models employ the metallicity profile and SFH of \citet{2006MNRAS.366..899N}.  The total disk mass in all of the models is $4.2\times10^{10}$M$_{\odot}$ between 0 kpc $<$R$<$15 kpc.
\item [2] Units of 10$^{11}$ stars between 2.5 kpc $<$R$<$15 kpc.
\item [3] The stellar number density at the solar neighbourhood in units of stars pc$^{-3}$.
\end{tablenotes}
\end{threeparttable}

\centering

\end{table}
\end{centering}

\clearpage

\begin{centering}
\begin{table}[!h]
\caption{Major events in Earth's History that inform the timescales for the emergence or re-emergence of complex life}
\resizebox {0.47\textwidth}{!}{
\centering\small
\begin{tabular}{cc}
\hline\hline
Event Name & Time of occurrence (Gya)\\
\hline
Rise of metazoan life&0.75\\
\hline
Formation of the ozone layer&2.3\\
\hline
Evidence of cyanobacteria&2.7\\
\hline
Formation of the Earth&4.55\\
\hline
\end{tabular}}
\centering
\label{ex:complex_life}
\end{table}
\end{centering}

\clearpage
\begin{centering}
\begin{table}[!h]
\caption{The main characteristics that are applied to each star in the model}
\resizebox {1\textwidth}{!}{
\centering\small
\begin{tabular}{lp{3.8in}}
\hline\hline
Stellar Property & Formula or Input Parameter\\
\hline
Initial Mass Function&Salpeter or Kroupa \\
\hline
Birth date&The SFH given by~\citet{2006MNRAS.366..899N}\\
\hline
Death date&After the main sequence lifetime expires (Equation~\ref{eqn:stellar_lifetime})\\
\hline
Position&Stellar number density distribution of~\citet{2006ima..book.....C} or \citet{2008ApJ...673..864J}\\
\hline
Metallicity&The metallicity distribution from~\citet{2006MNRAS.366..899N}\\
\hline
SNII&All stars with main sequence $M>8~M_\odot$~\citep{1984ApJ...277..361K}\\
\hline
SNIa&1\% of white dwarfs~\citep{2008ApJ...683L..25P}\\
\hline
Sterilization distance&Normalized to the average SNII sterilization distance of 8 pc~\citep{2003ApJ...585.1169G}\\
\hline
Having a habitable planet&A function of the planet-metallicity correlation~\citep{2005ApJ...622.1102F} and a model of solar system formation~\citep{2005ApJ...626.1045I}. A habitable planet has $0.1~M_\oplus<M_p<10~M_\oplus$, and exists in the HZ of its host star\\
\hline
Tidally-locked planet&Those with an appropriate mass, corresponding to the tidal-locking line of~\citet{1993Icar..101..108K}\\
\hline
Ozone reconstruction time&A value chosen uniformly in the range [0.4,2.25] Gyr\\
\hline
Animal re-evolution time&A value chosen uniformly in the range [0,1.55] Gyr\\
\hline
\end{tabular}}
\centering
\label{ex:table_summary}
\end{table}
\end{centering}

\end{document}